\documentclass[twocolumn,showpacs,preprintnumbers,amsmath,amssymb,floatfix]{revtex4}
\usepackage{bm}		
\usepackage{graphicx}	
\usepackage{dcolumn}	

\begin{document}

\title[Inside charged black holes I]{Inside charged black holes I. Baryons}

\author{Andrew J S Hamilton}
\email{Andrew.Hamilton@colorado.edu}
\homepage{http://casa.colorado.edu/~ajsh/}
\author{Scott E Pollack}
\email{pollacks@jilau1.colorado.edu}
\homepage{http://onehertz.colorado.edu/}
\affiliation{
JILA and Dept.\ Astrophysical \& Planetary Sciences,
Box 440, U. Colorado, Boulder CO 80309, USA
}

\newcommand{\dd}{d}
\newcommand{\gvec}{\bm{g}}
\newcommand{\pvec}{\bm{p}}
\newcommand{\gammavec}{\bm{\gamma}}
\newcommand{\xivec}{\bm{\xi}}
\newcommand{\ucoord}{\upsilon}
\newcommand{\utet}{u}
\newcommand{\partialvec}{\bm{\partial}}
\newcommand{\lnt}{\ln\!t}
\newcommand{\perpperp}{\perp\!\!\perp}
\newcommand{\Mc}{M_c}

\newcommand{\rnorm}{r}
\newcommand{\alphanorm}{\alpha}
\newcommand{\betanorm}{\beta}
\newcommand{\gammanorm}{\gamma}

\newcommand{\grarb}{\gprime}
\newcommand{\hrarb}{\hprime}
\newcommand{\rrarb}{\rprime}
\newcommand{\alphararb}{\alphaprime}
\newcommand{\betararb}{\betaprime}
\newcommand{\gammararb}{\gammaprime}
\newcommand{\lambdararb}{\lambdaprime}
\newcommand{\partialrarb}{\partialprime}

\newcommand{\rgen}{\rprimeprime}
\newcommand{\alphagen}{\alphaprimeprime}
\newcommand{\betagen}{\betaprimeprime}
\newcommand{\gammagen}{\gammaprimeprime}
\newcommand{\lambdagen}{\lambdaprimeprime}
\newcommand{\mugen}{\muprimeprime}
\newcommand{\nugen}{\nuprimeprime}
\newcommand{\partialgen}{\partialprimeprime}

\newcommand{\gtilde}{{\widetilde{g}}}
\newcommand{\htilde}{{\widetilde{h}}}
\newcommand{\rtilde}{{\widetilde{r}}}
\newcommand{\alphatilde}{\widetilde{\alpha}}
\newcommand{\betatilde}{\widetilde{\beta}}
\newcommand{\gammatilde}{\widetilde{\gamma}}
\newcommand{\lambdatilde}{\widetilde{\lambda}}

\newcommand{\gprime}{{g^\prime}}
\newcommand{\hprime}{{h^\prime}}
\newcommand{\rprime}{{r^\prime}}
\newcommand{\alphaprime}{{\alpha^\prime}}
\newcommand{\betaprime}{{\beta^\prime}}
\newcommand{\gammaprime}{{\gamma^\prime}}
\newcommand{\lambdaprime}{{\lambda^\prime}}
\newcommand{\muprime}{{\mu^\prime}}
\newcommand{\nuprime}{{\nu^\prime}}
\newcommand{\partialprime}{\partial^\prime}

\newcommand{\rprimeprime}{{r^{\prime\prime}}}
\newcommand{\alphaprimeprime}{{\alpha^{\prime\prime}}}
\newcommand{\betaprimeprime}{{\beta^{\prime\prime}}}
\newcommand{\gammaprimeprime}{{\gamma^{\prime\prime}}}
\newcommand{\lambdaprimeprime}{{\lambda^{\prime\prime}}}
\newcommand{\muprimeprime}{{\mu^{\prime\prime}}}
\newcommand{\nuprimeprime}{{\nu^{\prime\prime}}}
\newcommand{\partialprimeprime}{\partial^{\prime\prime}}

\hyphenpenalty=3000

\newcommand{\qmfig}{
    \begin{figure}
    \includegraphics[scale=.7]{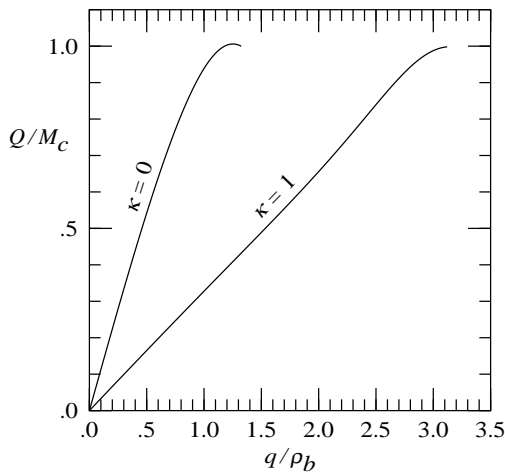}
    \caption[1]{
    \label{qm}
Charge-to-mass
$Q/\Mc$
of the black hole
as a function of the
charge-to-mass density
$q/\rho_b$
of baryons at the outer sonic point boundary,
for an accretion rate $\eta_s = 0.1$.
Two cases are shown,
one with zero conductivity, $\kappa = 0$,
the other with finite conductivity, $\kappa = 1$.
The charge-to-mass $Q/\Mc$ of the black hole
generally increases as the
charge-to-mass density
$q/\rho_b$
at the outer boundary is increased,
although in the zero conductivity case
the charge-to-mass $Q/\Mc$ does turn down slightly
at the largest charge-to-mass densities $q/\rho_b$.
The maximum charge-to-mass density $q/\rho_b$
is set by the condition that $\gamma^2 - \beta^2 > 0$.
    }
    \end{figure}
}

\newcommand{\varsSchwfig}{
    \begin{figure}
    \includegraphics[scale=.6]{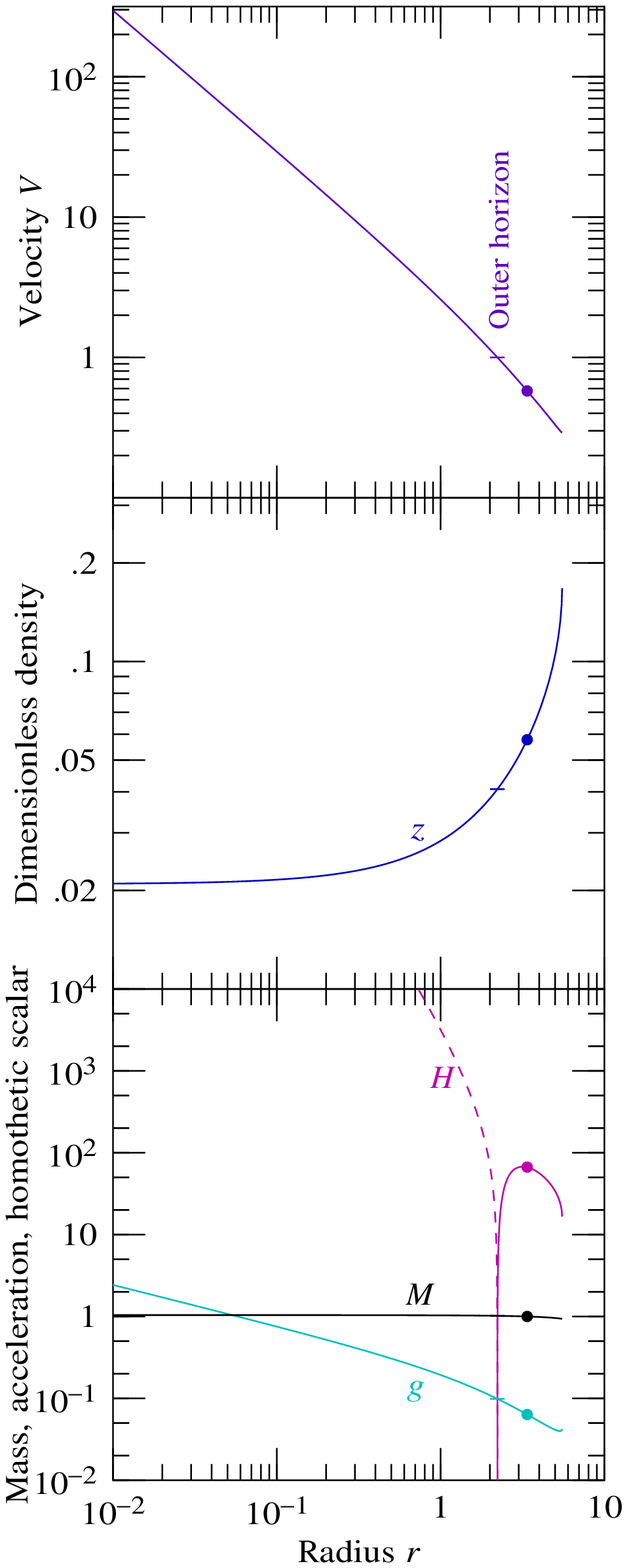}
    \caption[1]{
    \label{varsSchw}
(Color online)
The simplest model considered in this paper:
a black hole that accretes uncharged baryons,
at rate $\eta_s = 0.1$.
Quantities are plotted against radius $r$
in units where the charge-augmented mass at the outer sonic point is unity,
$\Mc = 1$.
Lines are dashed where quantities are negative.
Disks mark the outer sonic point,
where $V = \sqrt{w} \approx 0.577$,
at which the boundary conditions are set.
Short horizontal bars mark the horizon, where $V = 1$.
(Upper panel)
Proper velocity $V$ of the similarity frame
relative to the baryonic frame.
(Middle panel)
Dimensionless proper baryonic mass density
$z \equiv 4 \pi r^2 \rho_b$.
(Bottom panel)
Interior mass $M$,
proper acceleration $g$ experienced by the baryonic fluid,
and the homothetic scalar $H$, equation~(\protect\ref{H}).
    }
    \end{figure}
}

\newcommand{\penroseSchwfig}{
    \begin{figure}
    \includegraphics[scale=.9]{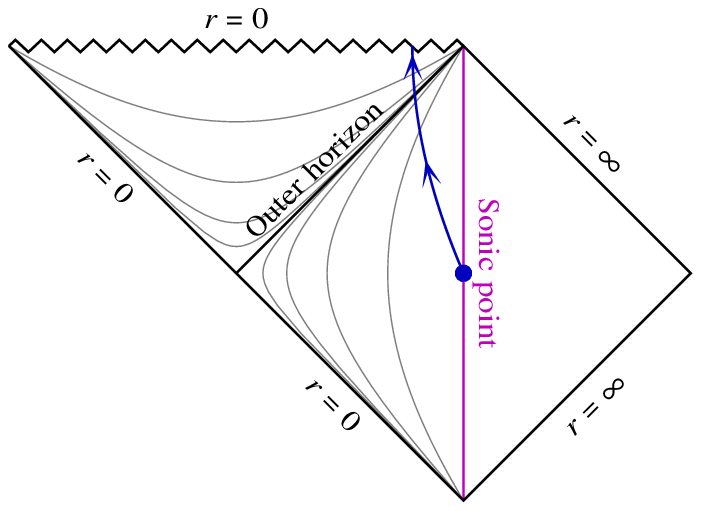}
    \caption[1]{
    \label{penroseSchw}
(Color online)
Penrose diagram
for the uncharged black hole
of Figure~\protect\ref{varsSchw}.
The arrowed line
represents schematically the worldline of a parcel of baryonic fluid.
The thin grey lines are lines of constant self-similar coordinate.
The self-similar solution extends some way outside the sonic point
outside the outer horizon,
but it does not continue to $r = \infty$;
the Penrose diagram as drawn presumes that the spacetime
joins to an asymptotically flat space as $r \rightarrow \infty$.
    }
    \end{figure}
}

\newcommand{\varsQMfig}{
    \begin{figure}
    \includegraphics[scale=.6]{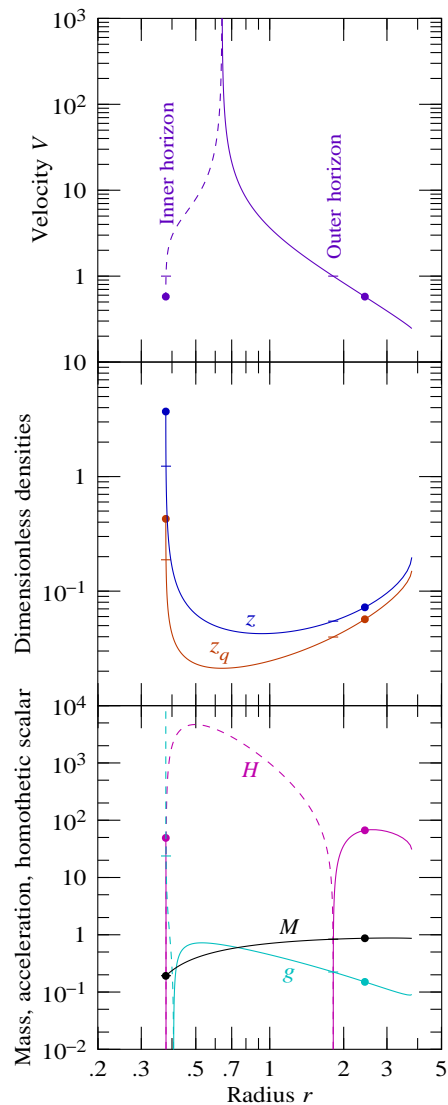}
    \caption[1]{
    \label{varsQM}
(Color online)
Similar to Figure~\protect\ref{varsSchw},
but for a black hole with charge-to-mass $Q/\Mc = 0.8$,
accreting baryons with zero conductivity, $\kappa = 0$.
Lines are dashed where quantities are negative.
Disks mark sonic points, where $V = \pm \sqrt{w}$.
Short horizontal bars mark horizons, where $V = \pm 1$.
(Upper panel)
Proper velocity $V$ of the similarity frame
relative to the baryonic frame.
The velocity passes from $V = + \infty$ to $- \infty$
as the charged baryonic fluid passes from ingoing to outgoing.
The baryonic fluid drops through the Cauchy horizon,
beyond which the solution terminates at an irregular sonic point.
(Middle panel)
Dimensionless proper baryonic mass and charge densities
$z \equiv 4 \pi r^2 \rho_b$
and
$z_q \equiv 4 \pi r^2 q$.
(Bottom panel)
Interior mass $M$,
proper acceleration $g$,
and the homothetic scalar $H$.
    }
    \end{figure}
}

\newcommand{\shkQMfig}{
    \begin{figure}
    \includegraphics[scale=.6]{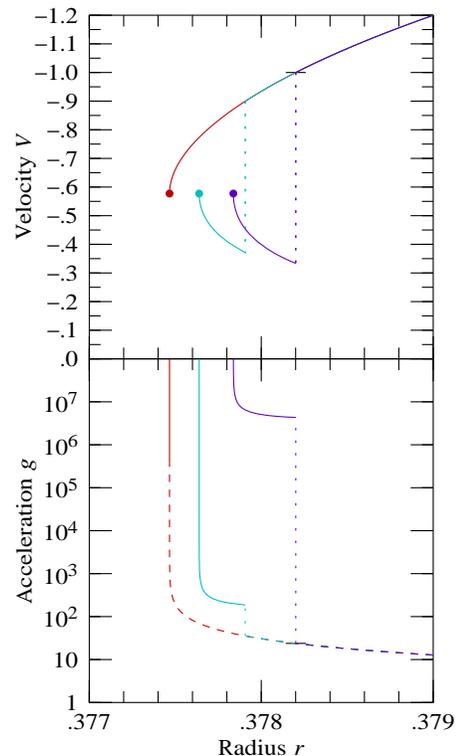}
    \caption[1]{
    \label{shkQM}
(Color online)
Putative continuations of the
similarity solution
inside the Cauchy horizon
for the model shown in Figure~\protect\ref{varsQM},
which has $Q/\Mc = 0.8$ and zero conductivity.
Three cases are shown,
ranging the gamut
from an extremely weak shock to an extremely strong shock.
In all cases
the shocked fluid continues downstream for a short distance
before terminating at an irregular sonic point, marked by a disk,
where the proper acceleration diverges
and the similarity solution cannot be continued.
Dotted lines connect pre-shock and post-shock quantities.
(Top panel)
Proper velocity $V$ of the similarity frame
relative to the baryonic frame.
The Cauchy horizon, where $V = -1$, is marked by a short horizontal bar.
A shock decelerates the baryonic fluid from supersonic
to subsonic,
through the sound speed where $V = - \sqrt{w} \approx 0.577$.
(Bottom panel)
Proper acceleration $g$ experienced by the baryonic fluid
in the vicinity of each of the three putative shocks.
The acceleration changes from outward (negative, dashed lines)
in the pre-shock fluid to inward (positive, continuous lines)
in the post-shock fluid.
    }
    \end{figure}
}

\newcommand{\penroseQMfig}{
    \begin{figure}
    \includegraphics[scale=.9]{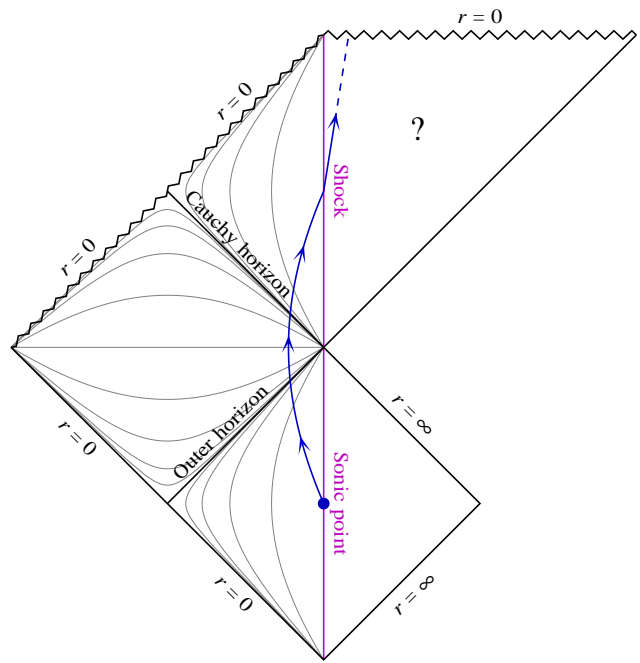}
    \caption[1]{
    \label{penroseQM}
(Color online)
Tentative Penrose diagram
for a charged black hole,
such as that shown in Figures~\protect\ref{varsQM} and \protect\ref{shkQM},
in which the baryons drop through a Cauchy horizon
and undergo a shock before falling to a central spacelike singularity.
The arrowed line
represents schematically the worldline of a parcel of baryonic fluid.
The thin grey lines are lines of constant self-similar coordinate.
The self-similar solution extends some way outside the sonic point
outside the outer horizon,
but it does not continue to $r = \infty$.
The Penrose diagram as drawn presumes that the spacetime
joins to an asymptotically flat space as $r \rightarrow \infty$.
Similarly,
the self-similar solution extends some way inside the shock
inside the Cauchy horizon,
but it does not continue to a singularity,
but rather terminates at an irregular sonic point.
The Penrose diagram as drawn presumes that the spacetime
terminates at a spacelike singularity at $r = 0$.
A question mark emphasizes the fact that
the similarity solution leaves undetermined how
the spacetime continues (or not) to the singularity.
    }
    \end{figure}
}

\newcommand{\velfig}{
    \begin{figure*}
    \includegraphics[scale=.9]{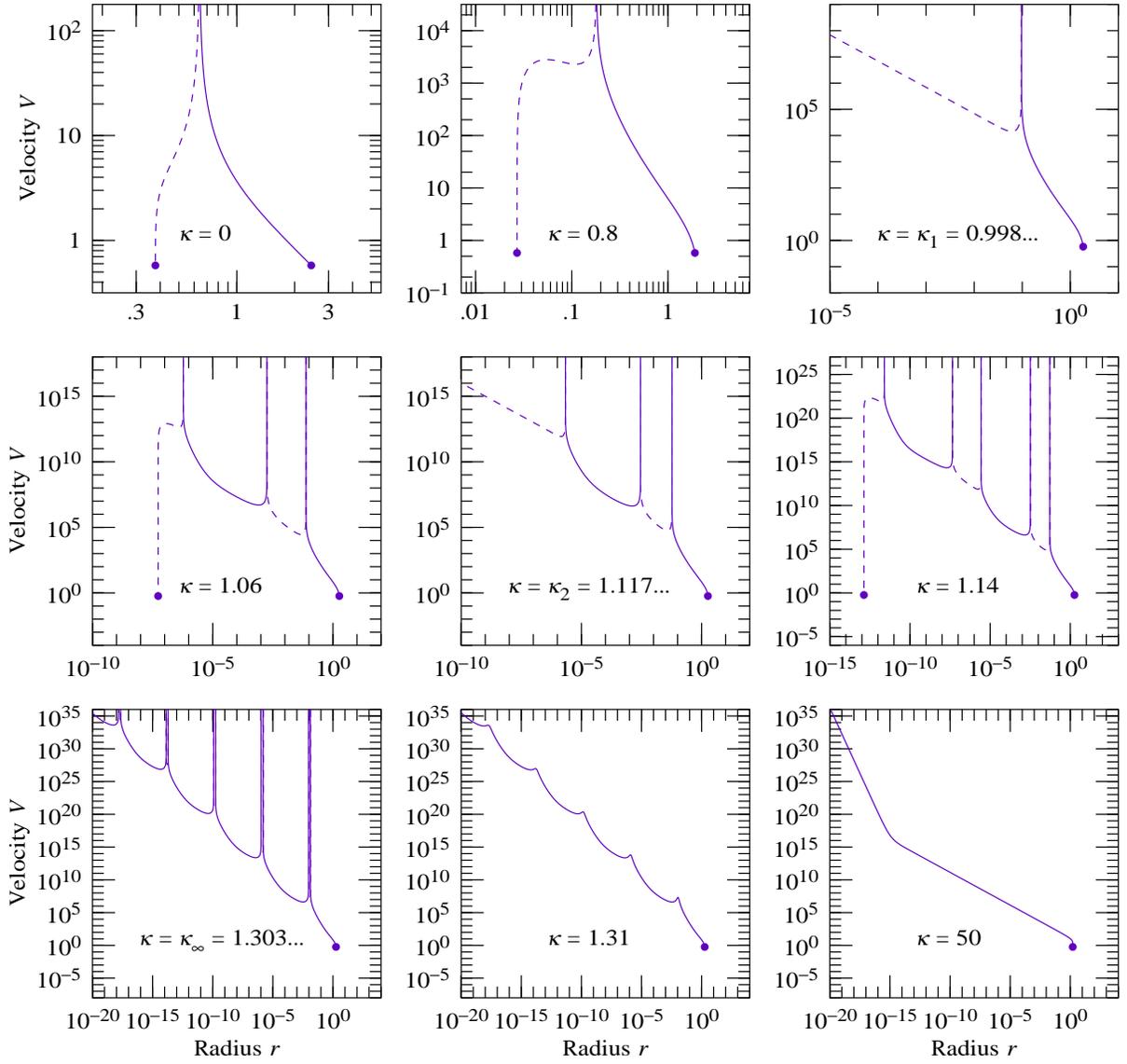}
    \caption[1]{
    \label{vel}
Velocity $V$ versus radius $r$
for pure bayonic models
with various conductivities $\kappa$,
increasing from zero at top left
to a large value somewhat less than
the maximum conductivity $\kappa_{\max} \approx 67.0$ at bottom right.
At conductivities less than the continuum threshold $\kappa_\infty$,
the top six panels,
the baryonic fluid drops through the Cauchy horizon
unless the conductivity equals one of
an infinite spectrum of discrete values $\kappa_n$,
in which case the fluid falls to a singularity at zero radius.
The first two cases of the discrete spectrum,
$\kappa_1$ and $\kappa_2$, are shown.
The velocity $V$ passes through $\pm \infty$
where the baryonic fluid transitions between ingoing ($V > 0$, continuous lines)
and outgoing ($V < 0$, dashed lines).
Solutions between values of the discrete spectrum $\kappa_n$
are characterized by the number of times that the baryonic fluid
cycles between ingoing and outgoing.
At the continuum threshold $\kappa_\infty$,
the bottom left panel,
the solution exhibits discrete self-similarity,
the baryonic fluid cycling repeatedly between ingoing and outgoing.
At and above the continuum threshold $\kappa_\infty$,
the bottom three panels,
all solutions fall to a central singularity.
Disks mark sonic points, where $V = \pm \sqrt{w} \approx \pm 0.577$.
Horizons occur where $V = \pm 1$.
    }
    \end{figure*}
}

\newcommand{\velQMonefig}{
    \begin{figure}
    \includegraphics[scale=.6]{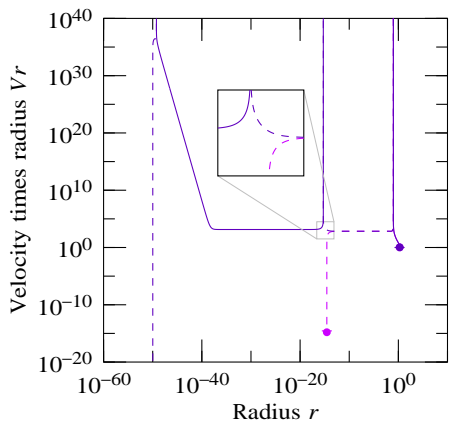}
    \caption[1]{
    \label{velQM1}
(Color online)
Velocity times radius, $V r$,
for two models with conductivities
$\kappa = \kappa_1 \pm \epsilon$
just either side of the value $\kappa_1 \approx 0.998$
that separates solutions that cycle once versus twice
between ingoing ($V > 0$, continuous lines)
and outgoing ($V < 0$, dashed lines).
The inset shows a detail of how the velocities in the two models
peel off from the critical solution,
which itself would continue on the outgoing path
as a horizontal line to zero radius.
    }
    \end{figure}
}

\newcommand{\varsQMkfig}{
    \begin{figure}
    \includegraphics[scale=.6]{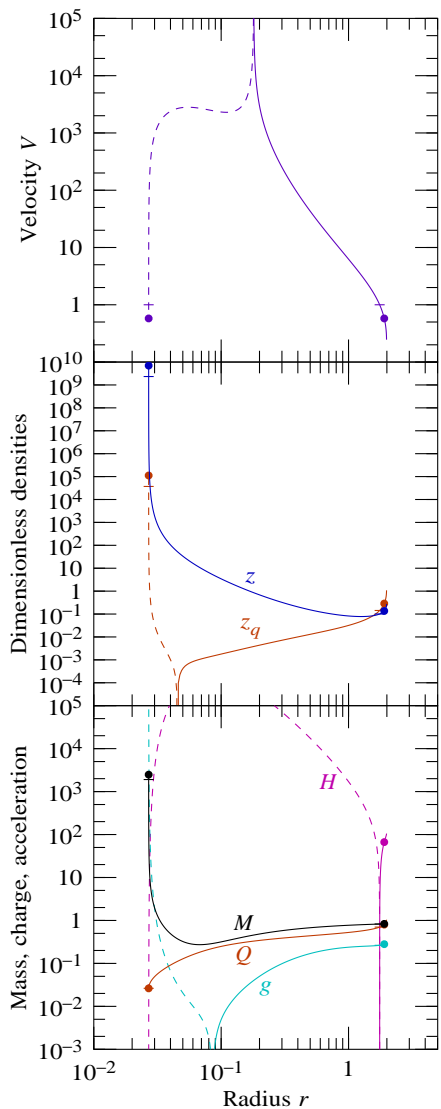}
    \caption[1]{
    \label{varsQMk}
(Color online)
Similar to Figures~\protect\ref{varsSchw} and \protect\ref{varsQM}.
Model with conductivity $\kappa = 0.8$,
less than the first critical conductivity $\kappa_1$.
The baryonic fluid drops through the Cauchy horizon.
    }
    \end{figure}
}

\newcommand{\varsQMcfig}{
    \begin{figure}
    \includegraphics[scale=.6]{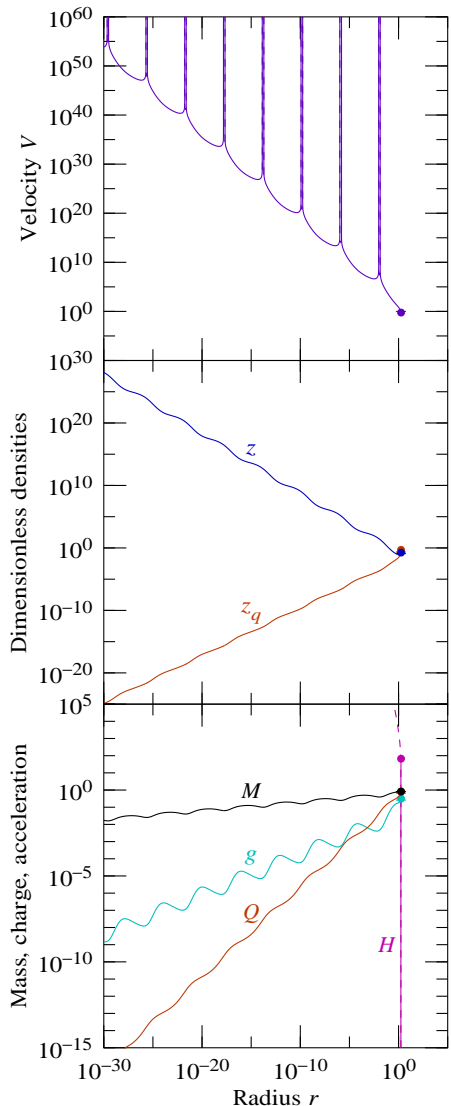}
    \caption[1]{
    \label{varsQMc}
(Color online)
Similar to Figures~\protect\ref{varsSchw}, \protect\ref{varsQM}.
Model with conductivity at the continuum threshold conductivity
$\kappa = \kappa_\infty$.
The model displays discrete self-similarity.
    }
    \end{figure}
}

\newcommand{\scalingQMcfig}{
    \begin{figure}
    \includegraphics[scale=.6]{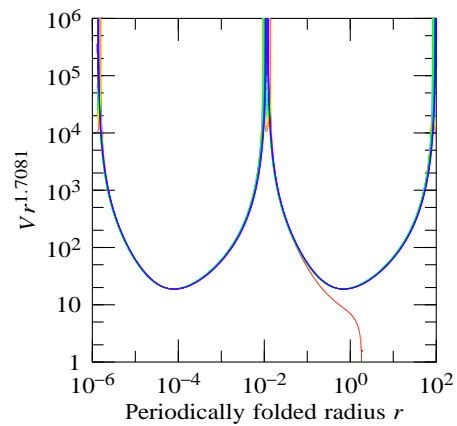}
    \caption[1]{
    \label{scalingQMc}
(Color online)
Discrete self-similarity at the continuum threshold conductivity
$\kappa = \kappa_\infty$.
The horizontal axis is the radius $r$ folded
over two periods, a period being $3.9463$ decades.
The vertical axis is the velocity $V$
multiplied by a power of radius,
$r^{1.7081}$.
The solution as shown passes through 31 periods,
covering a total of about 120 decades of radius.
    }
    \end{figure}
}

\newcommand{\appearfig}{
    \begin{figure*}
    \includegraphics[scale=.525]{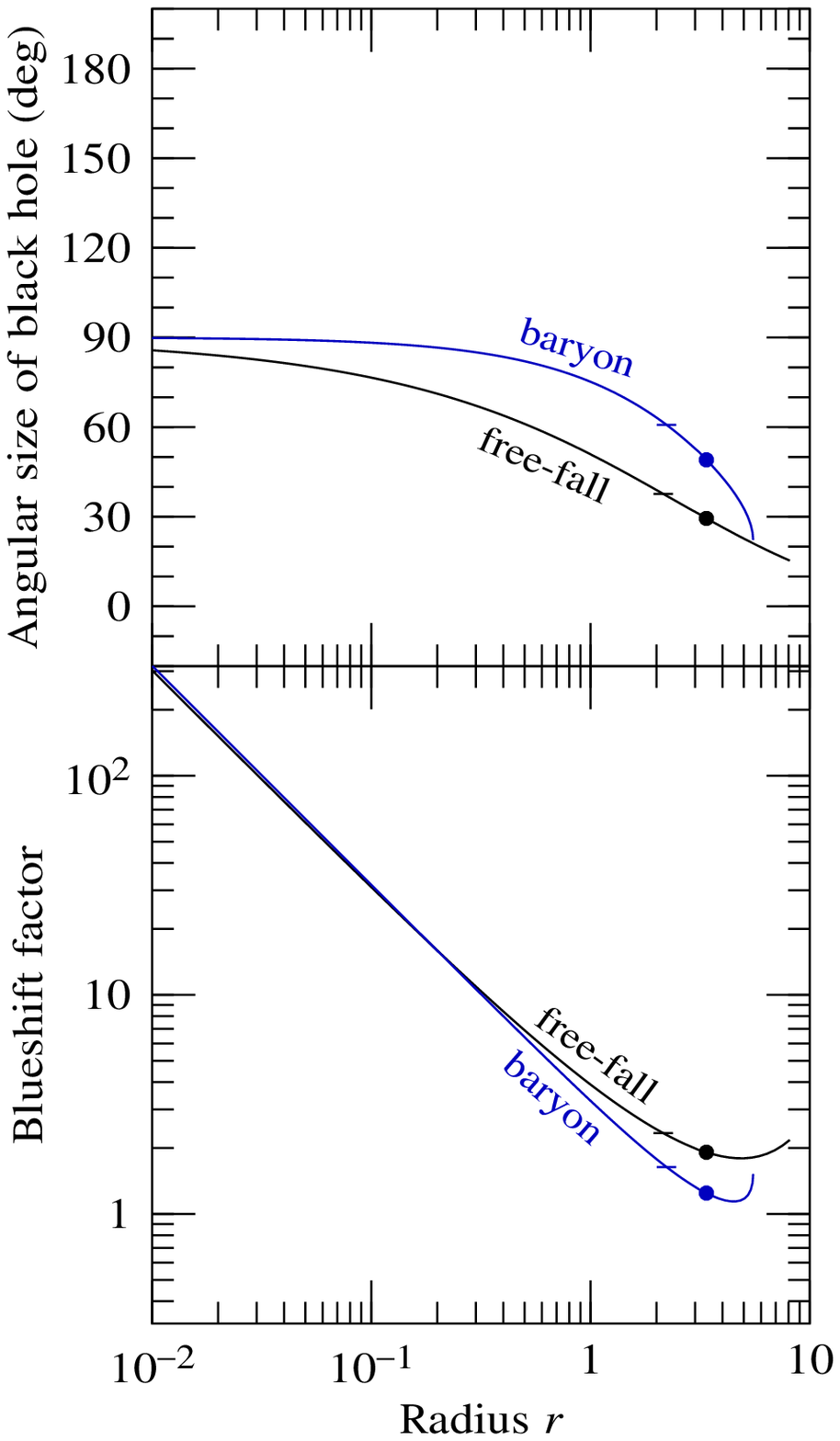}
    \includegraphics[scale=.525]{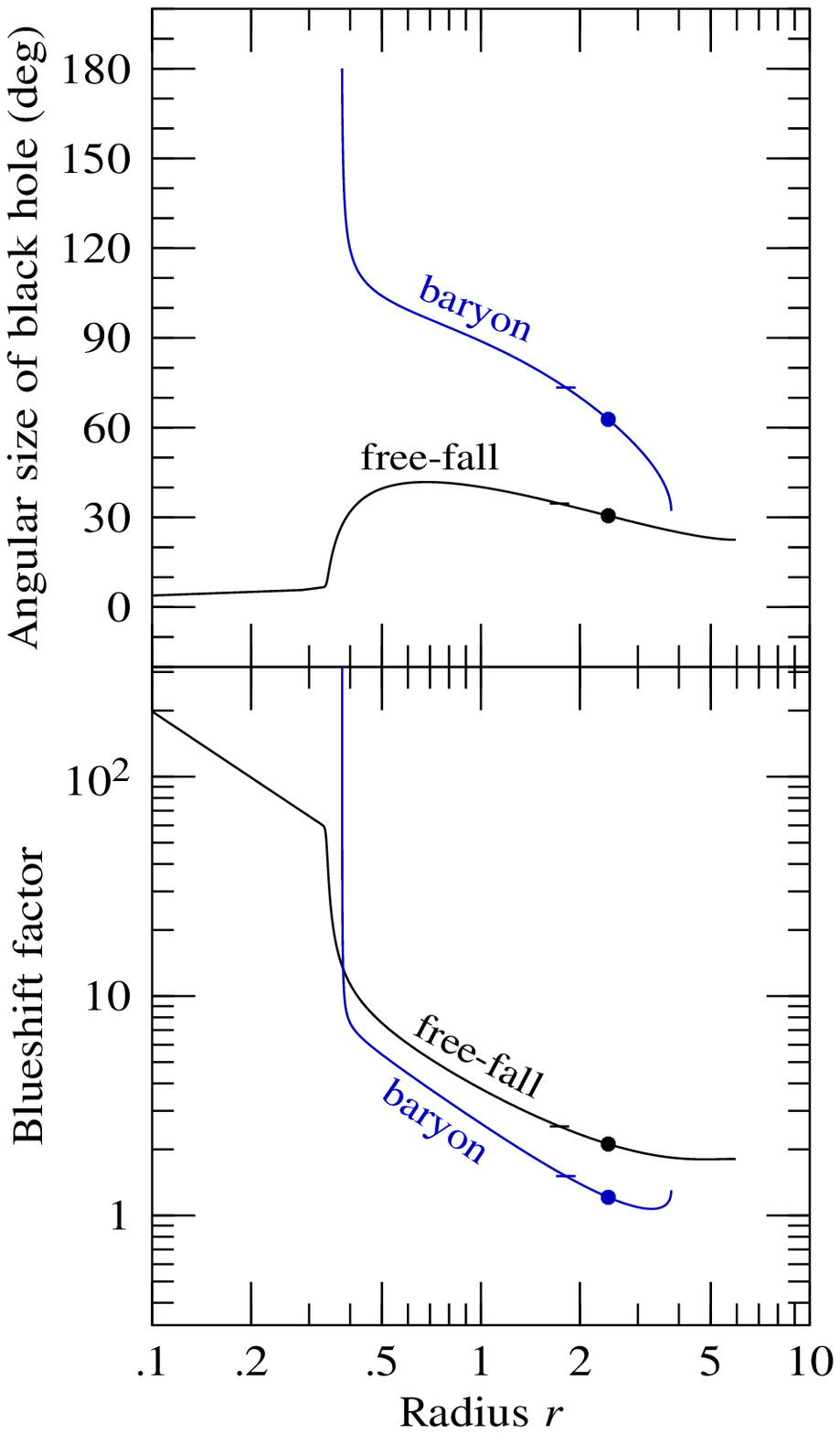}
    \includegraphics[scale=.525]{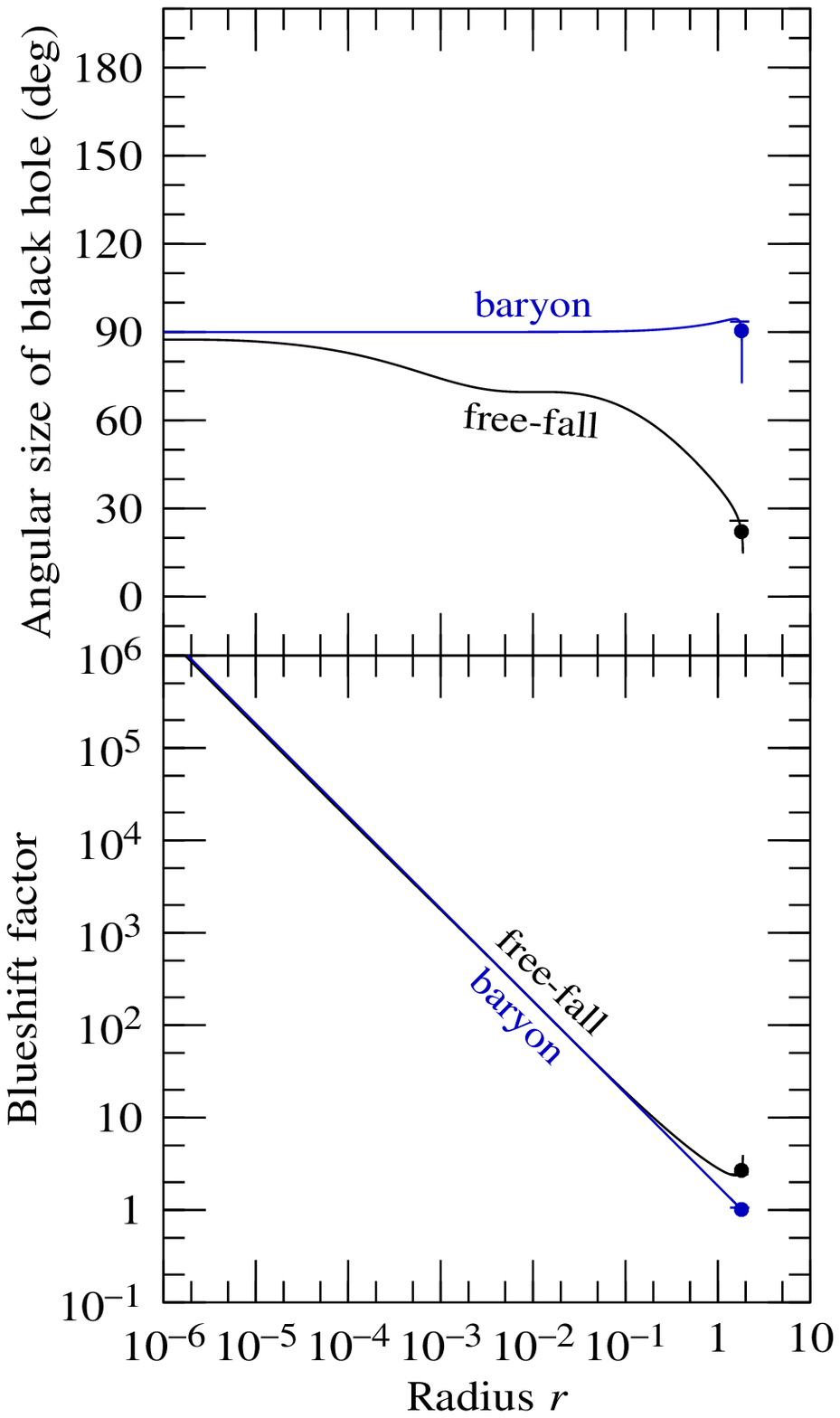}
    \caption[1]{
    \label{appear}
(Color online)
Angular size $\chi_\textrm{ph}$ of the black hole,
equation~(\protect\ref{chiph}),
and the blueshift of photons at the edge of the black hole,
equation~(\protect\ref{blueph}),
perceived by an observer
in either the baryonic frame or a radial free-fall frame,
for three models.
The models shown are:
(left) the uncharged baryonic model
from Figure~\protect\ref{varsSchw};
(middle) the charged non-conducting baryonic model
from Figure~\protect\ref{varsQM};
(right) the discretely self-similar
charged baryonic model with
$\kappa = \kappa_\infty$,
from Figure~\protect\ref{varsQMc}.
Disks mark the outer sonic point, where $V = \sqrt{w} \approx 0.577$.
Horizontal bars mark the outer horizon, where $V = 1$.
Light from the outside universe is visible only from outside
the Cauchy horizon,
so lines terminate infinitesimally outside the Cauchy horizon
even in models in which the baryons drop inside the Cauchy horizon
(middle model).
    }
    \end{figure*}
}

\newcommand{\rSchwtab}{
    \begin{table}[htb]
    \caption{
    \label{rSchw}
Radii of events in an uncharged black hole accreting baryons at rate
$\eta_s = 0.1$,
compared to corresponding radii
in the Schwarzschild (Schw) solution.
}
    \begin{ruledtabular}
    \begin{tabular}{lll}
Event & Model & Schw \\
\hline
Sonic point & 3.377 & 3 \\
Photon sphere & 3.176 & 3 \\
Outer horizon & 2.222 & 2
    \end{tabular}
    \end{ruledtabular}
    \end{table}
}

\newcommand{\rQMtab}{
    \begin{table}[htb]
    \caption{
    \label{rQM}
Radii of events
for the model with
accretion rate $\eta_s = 0.1$,
charge-to-mass $Q/\Mc = 0.8$,
and zero conductivity,
compared to corresponding radii
in the Reissner-Nordstr\"om solution with
the same charge-to-mass
$Q/\Mc = 0.8$.
}
    \begin{ruledtabular}
    \begin{tabular}{lll}
Event & Model & RN \\
\hline
Photon sphere & 2.439 & 2.485 \\
Sonic point & 2.439 & 2.122 \\
Outer horizon & 1.813 & 1.6 \\
$(\beta,\gamma)$ becomes timelike & 1.643 & 1.6 \\
Charged particles become outgoing & 0.639 & 0.64 \\
$\gamma$ becomes negative & 0.614 & 0.64 \\
Inner horizon & 0.3782 & 0.4 \\
$(\beta,\gamma)$ becomes spacelike & -- & 0.4 \\
Sonic point & 0.3775 & 0.3994
    \end{tabular}
    \end{ruledtabular}
    \end{table}
}

\newcommand{\kappatab}{
    \begin{table}[hbt]
    \caption{
    \label{kappa}
Spectrum of critical conductivities
for a black hole
with
accretion rate $\eta_s = 0.1$
and
charge-to-mass $Q/\Mc = 0.8$.
}
    \begin{ruledtabular}
    \begin{tabular}{ll}
$\kappa_1$ & 0.9982939540115 \\
$\kappa_2$ & 1.1177942541104 \\
$\kappa_3$ & 1.1654295843302 \\
$\kappa_4$ & 1.1903424212409 \\
$\kappa_\infty$ & 1.30382 \\
$\kappa_{\max}$ & 66.998112873734 \\
    \end{tabular}
    \end{ruledtabular}
    \end{table}
}

\begin{abstract}
An extensive investigation is made
of the interior structure of self-similar accreting charged black holes.
In this, the first of two papers,
the black hole is assumed to accrete a charged, electrically conducting,
relativistic baryonic fluid.
The mass and charge of the black hole are generated self-consistently
by the accreted material.
The accreted baryonic fluid undergoes one of two possible fates:
either
it plunges directly to the spacelike singularity at zero radius,
or else
it drops through the Cauchy horizon.
The baryons fall directly to the singularity if the conductivity
either exceeds a certain continuum threshold $\kappa_\infty$,
or else equals one of an infinite spectrum $\kappa_n$ of discrete values.
Between the discrete values $\kappa_n$,
the solution is characterized by the number of times that
the baryonic fluid cycles between ingoing and outgoing.
If the conductivity is at the continuum threshold $\kappa_\infty$,
then the solution cycles repeatedly between ingoing and outgoing,
displaying a discrete self-similarity
reminiscent of that observed in critical collapse.
Below the continuum threshold $\kappa_\infty$,
and except at the discrete values $\kappa_n$,
the baryonic fluid drops through the Cauchy horizon,
and in this case undergoes a shock,
downstream of which the solution terminates at an irregular sonic point
where the proper acceleration diverges,
and there is no consistent self-similar continuation to zero radius.
As far as the solution can be followed inside the Cauchy horizon,
the radial direction is timelike.
If the radial direction remains timelike to zero radius
(which cannot be confirmed because the self-similar solutions terminate),
then there is presumably a spacelike singularity at zero radius
inside the Cauchy horizon,
which is distinctly different from the vacuum (Reissner-Nordstr\"om) solution
for a charged black hole.
\end{abstract}

\pacs{04.20.-q}	

\date{\today}

\maketitle

\section{Introduction}

What really happens inside black holes?
Despite substantial progress,
particularly in the last decade and a half
since the landmark papers by
Poisson \& Israel (1990) \cite{PI90}
on mass inflation at the Cauchy horizon,
by Ori \& Piran (1990) \cite{OP90}
on similarity solutions in general relativistic collapse,
and by Choptuik (1993) \cite{Choptuik93}
on critical collapse
and discrete self-similarity,
the answer to this question remains unresolved,
e.g.\ \cite{Berger02,Gundlach03,Novikov03,Dafermos03,Dafermos04} and references therein.

It is well known that the vacuum solutions for charged
(Reissner-Nordstr\"om)
and rotating
(Kerr-Newman)
black holes are not physically consistent
as endpoints of realistic gravitational collapse,
because their cores are gravitationally repulsive.
For a charged black hole,
the gravitational repulsion comes from the negative radial pressure
(radial tension) of the electric field,
while for a rotating black hole,
the gravitational repulsion comes from the centrifugal force.
The gravitational repulsion causes accreted material
to tend to pile up in the subluminal region inside the inner horizon
of the vacuum black hole,
rather than falling on to the singularity,
contradicting the proposition that
the black hole is empty outside the singularity.
Only in the unique case of an
uncharged, non-rotating black hole (Schwarzschild)
is the vacuum solution a consistent endpoint
of realistic gravitational collapse.

The present paper follows a well-traveled trail of exploration
into the interiors of black holes,
by considering only spherically symmetric self-similar solutions,
and imagining that it might be reasonable to regard
charge as a surrogate for rotation,
e.g.\ \cite{Dafermos04}.
General relativistic spherically symmetric self-similar solutions
were first considered by
Cahill \& Taub (1971) \cite{CT71},
and were first applied to the problem
of general relativistic gravitational collapse by
Ori and Piran (1990) \cite{OP90},
following earlier work
on general relativistic self-similar solutions in cosmology.
Since then there have been many investigations;
see especially the reviews
\cite{CC99,Harada03} and references therein;
and more recently
\cite{MTW03,HM03,MH04,PW04,WWW04,CsDVdRW04,SA04}.

The driving goal of this paper
is not so much to study the formation of a black hole by
gravitational collapse,
but rather to explore the interior structure of black holes
after their formation.
We have in mind the situation of a realistic astronomical black hole,
perhaps stellar-sized, perhaps supermassive,
which is being fed by accretion of baryonic matter.
The accretion rates in astronomical black holes are typically tiny,
in the sense that the accretion timescale, which might be millions to
billions of years, is vast compared to the characteristic timescale
of the black hole, which might be milliseconds to hours.
Thus it is to be expected that the superluminal region outside
the inner horizon of a black hole should closely approximate
the vacuum solution.
The interesting question is where and how the interior structure
of the black hole deviates from the vacuum solution.

In a seminal paper,
Poisson \& Israel (1990)
\cite{PI90}
showed that if ingoing and outgoing fluids are allowed to pass through
each other inside a charged black hole,
then the generic consequence is
`mass inflation'
as the counter-streaming fluids approach an inner horizon.
During mass inflation,
the interior mass, the so-called Misner-Sharp mass \cite{MS64},
a gauge-invariant scalar quantity,
exponentiates to enormous values.
The phenomenon of mass inflation has been confirmed
analytically and numerically
in many papers
\cite{Ori91,BDIM94,BS95,BO98,Burko97,Burko03,Dafermos04}.

Physically,
mass inflation is a consequence of the fact that,
if ingoing and outgoing fluids drop through an inner horizon,
then they must necessarily pass into causally disconnected
parts of spacetime.
The only way this can happen is for the ingoing and outgoing fluids
to exceed the speed of light relative to each other,
which is impossible.
As the counter-streaming fluids race through each other at ever closer to
the speed of light in their attempt to drop through the inner horizon,
they generate a large radial pressure.
The growing radial pressure amplifies the gravitational force
that accelerates the ingoing and outgoing fluids through each other,
which results in mass inflation
(see \S{}V of \cite{Paper2} for a more comprehensive discussion).



Analytic and numerical work
on spherically symmetric collapse
has commonly modeled the fluid accreted by the black hole
as a massless scalar field,
usually taken to be uncharged
\cite{Christodoulou86,GP87,GG93,BS95,Brady95,BO98,Burko97,Burko99,
HO01,Burko02a,Burko03,MG03,Dafermos04,HKN05},
but sometimes charged
\cite{HP96,HP98b,HP99,SP01,OP03,Dafermos03}.
A key property of a massless scalar field
is that it allows waves to counter-stream relativistically through each other,
which allows the phenomenon of mass inflation on the Cauchy horizon to occur.
In the present paper and its companion, hereafter Paper~2 \cite{Paper2},
we choose to adopt a somewhat different point of view,
treating it as a matter of physics
as to whether or not to allow relativistic counter-streaming,
rather than assuming a fluid that has that property built in.

In this paper,
we assume that the black hole is fed in a self-similar fashion
with a charged, conducting, baryonic fluid
having a relativistic equation of state
$p_b / \rho_b = 1/3$.
Although the fluid is electrically conducting,
so that oppositely charged fluids do drift through each other,
this proves not enough to permit mass inflation.
The charged baryonic fluid is electrically repelled
by the charge of the black hole,
generated self-consistently by the charge of the accreted matter,
and if the conductivity is small enough,
then the baryonic fluid naturally becomes outgoing,
and can drop through the outgoing inner horizon, the Cauchy horizon.

The electrical conductivity of the charged baryonic fluid is treated
in this paper as an adjustable parameter.
If the conductivity is set at a realistic level,
then the black hole accretes only if the charge-to-mass ratio
is set to a tiny value.
This is of course a consequence of the huge
charge-to-mass ratio of individual protons and electrons,
$e / m_\textrm{p} \approx 10^{18}$,
where $e$ is the dimensionless charge of the proton or electron,
the square root of the fine-structure constant,
and $m_\textrm{p}$ is the proton mass in units of the Planck mass.
However,
electrical conduction in a charged black hole
is analogous to angular momentum transport in a rotating black hole,
and angular momentum transport is a much weaker process
than electrical conduction
(if angular momentum transport were as strong as electrical conduction,
then accretion disks would shed angular momentum
as quickly as they shed charge,
and accretion disks would not rotate).
In the interests of regarding
charge as a surrogate for angular momentum,
it makes sense to investigate how the conductivity
affects the internal structure of the black hole,
with the conductivity being greatly suppressed compared
to any realistic conductivity,
but nevertheless possibly consistent with what might be a reasonable rate
for the analogous angular momentum transport in a rotating black hole.

In Paper~2 \cite{Paper2}
we allow the black hole to accrete,
in addition to the charged baryonic fluid,
a pressureless neutral `dark matter' fluid.
The point of introducing dark matter
is specifically to permit mass inflation,
caused by relativistic counter-streaming between outgoing baryons
and ingoing dark matter near the inner horizon.
We defer further discussion of mass inflation to Paper~2.

Previous investigations of self-similar solutions in gravitational collapse
have commonly required that the solutions be regular at the origin,
where the black hole first forms
\cite{OP90,GNU98,CC99,Harada03}.
The present paper does not impose this requirement from the outset,
but rather establishes boundary conditions outside the black hole,
and integrates the equations inward.
The outer boundary is set at the sonic point
where the infalling baryonic fluid is assumed to transition
smoothly from subsonic to supersonic velocity.
This point of view seems natural,
since information can only propagate inwards inside the black hole.


A final difference between the present paper
and earlier works is mathematical rather than physical.
In this paper and its companion
we choose to use the
tetrad formalism,
e.g.\ \cite{Chandrasekhar,BB03,GG05}.
While it is to some extent a matter of taste what gauge
or formalism to adopt,
physical results being independent of such choices,
the tetrad formalism does reveal the physics in an elegant and transparent way.
We owe much to the opus by Lasenby et al.\ \cite{LDG98},
which showed us how to proceed at an early stage of this project.
In particular, the gauge choices adopted here, equation~(\ref{gauge}),
are those suggested by \cite{LDG98}.

The structure of this paper is as follows.
Section~\ref{Equations}
presents the general relativistic equations governing
the interior and exterior structure of a spherically symmetric black hole
accreting a charged, electrically conducting fluid of baryons.
Section~\ref{Similaritysolutions}
introduces the hypothesis of self-similarity,
and sets out the equations that follow from that hypothesis.
Section~\ref{Resultsbaryons}
gives results for black holes that accrete only baryons.
Section~\ref{Appearance}
addresses the question of what it would actually look like if
you fell into one of the black holes described in this paper.
Finally,
section~\ref{Summary}
summarizes the findings of this paper.

\section{Equations}
\label{Equations}

This section
presents the general relativistic equations governing
a spherically symmetric black hole
accreting a charged, electrically conducting fluid of baryons.

\subsection{Tetrad formalism}
\label{tetradformalism}

Let $x^\mu$ denote a system of spacetime coordinates,
and let $\gvec_\mu$ denote the basis of tangent vectors
in that coordinate system.
By definition,
the scalar products of the tangent vectors constitute the metric $g_{\mu\nu}$
\begin{equation}
\label{gg}
  \gvec_\mu \cdot \gvec_\nu = g_{\mu\nu}
  \ .
\end{equation}
In the orthonormal tetrad formalism,
a set of locally inertial frames is erected at each point of the spacetime.
The locally inertial frame at each point has axes
$\gammavec_m$, the tetrad,
whose dot products form, by construction, the Minkowski metric
$\eta_{mn}$
\begin{equation}
  \gammavec_m \cdot \gammavec_n = \eta_{mn}
  \ .
\end{equation}
In this paper
dummy latin indices signify locally inertial, or tetrad, frames,
while dummy greek indices signify curved coordinate frames.
The axes $\gammavec_m$ of the locally inertial frames are related to
the tangent vectors $\gvec_\mu$
by the vierbein
${e_m}^\mu$
and its inverse
${e^m}_\mu$
\begin{equation}
  \gammavec_m = {e_m}^\mu \gvec_\mu
  \ , \quad
  \gvec_\mu = {e^m}_\mu \gammavec_m
  \ .
\end{equation}
The vierbein provide the means of transforming between
the tetrad components
and coordinate components of any 4-vector or tensor object.
For example,
the tetrad components $p^m$
and the coordinate components $p^\mu$ of the 4-vector
$\pvec \equiv p^\mu \gvec_\mu = p^m \gammavec_m$
are related by
$p^m = {e^m}_\mu p^\mu$.
Tetrad components are raised, lowered, and contracted
with the Minkowski metric $\eta_{mn}$,
while coordinate components are raised, lowered, and contracted
with the coordinate metric $g_{\mu\nu}$.

The most general form of the vierbein
${e_m}^\mu$
consistent with spherical symmetry is
\cite{Robertson},
in Cartesian coordinates,
\begin{eqnarray}
&&
  {e_0}^0 = \alpha
  \ , \quad
  {e_i}^0 = \nu \, \hat x_i
  \ , \quad
  {e_0}^i = \beta \, \hat x_i
  \ ,
\nonumber
\\
\label{genvierbein}
&&
  {e_i}^j
  =
  \gamma \, \hat x_i \hat x_j
  + \lambda \, (\delta_{ij} - \hat x_i \hat x_j)
  + \mu \, \varepsilon_{ijk} \hat x_k
\end{eqnarray}
where
$0$ is the time index,
$i,j,k$ are taken to run over spatial indices $1,2,3$,
and
$\varepsilon_{ijk}$
is
the completely antisymmetric flat space spatial 3-tensor.
Appendix~\ref{generaltetrad}
gives results for the general case,
but here it is convenient immediately to make the gauge choices
\cite{LDG98}
\begin{equation}
\label{gauge}
  \nu = 0
  \ , \quad
  \lambda = 1
  \ , \quad
  \mu = 0
\end{equation}
so that the vierbein
${e_m}^\mu$
simplify to
\begin{eqnarray}
&&
  {e_0}^0 = \alpha
  \ , \quad
  {e_i}^0 = 0
  \ , \quad
  {e_0}^i = \beta \, \hat x_i
  \ ,
\nonumber
\\
\label{vierbein}
&&
  {e_i}^j
  =
  \gamma \, \hat x_i \hat x_j
  + (\delta_{ij} - \hat x_i \hat x_j)
  \ .
\end{eqnarray}
The gauge choices
$\nu = 0$ and $\mu = 0$
can be effected
by starting from a general vierbein of the form~(\ref{genvierbein}),
and respectively boosting and rotating the tetrad frame appropriately at each point.
The gauge choice
$\lambda = 1$
then corresponds to scaling the radial coordinate $r$
so that the proper circumference of a circle of radius $r$ is $2\pi r$,
a traditional and natural choice.
The inverse vierbein ${e^m}_\mu$
corresponding to the vierbein
of equations~(\ref{vierbein}) is
\begin{eqnarray}
\label{inversevierbein}
&&
  {e^0}_0
  =
  {1 \over \alpha}
  \ , \quad
  {e^0}_i
  =
  0
  \ , \quad
  {e^i}_0
  =
  - {\beta \over \alpha \gamma} \hat x_i
  \ ,
\nonumber
\\
&&
  {e^j}_i
  =
  {1 \over \gamma} \hat x_i \hat x_j
  + (\delta_{ij} - \hat x_i \hat x_j)
  \ .
\end{eqnarray}

Directed derivatives $\partial_m$ are defined
to be the spacetime derivatives along the axes $\gammavec_m$
of the tetrad frame:
\begin{equation}
\label{directedderivative}
  \partial_m
  \equiv
  \gammavec_m \cdot \partialvec
  =
  {e_m}^\mu {\partial \over \partial x^\mu}
\end{equation}
where
$\partialvec = \gvec^\mu \partial / \partial x^\mu = g^{\mu\nu} \gvec_\nu \partial / \partial x^\mu$
is the invariant spacetime vector derivative.
The directed derivatives $\partial_m$
depend only on the choice of tetrad frame,
and are independent of the choice of coordinate system.
Unlike the coordinate partial derivatives $\partial / \partial x^\mu$,
the directed derivatives $\partial_m$ do not commute.
For the present paper, the most important directed derivatives
are the directed time derivative $\partial_t$
and the directed radial derivative $\partial_r$,
which for the vierbein of equations~(\ref{vierbein}) are
\begin{equation}
\label{directedderivativetr}
  \partial_t
  =
  \alpha {\partial \over \partial t} + \beta {\partial \over \partial r}
  \ , \quad
  \partial_r
  =
  \gamma {\partial \over \partial r}
  \ .
\end{equation}
The directed time derivative
$\partial_t$
is the proper time derivative
(commonly written $\dd / \dd \tau$)
along the worldline of an object
instantaneously at rest in the tetrad frame.
The coordinate 4-velocity
$\ucoord^\mu \equiv \partial_t x^\mu$
of an object instantaneously at rest in the tetrad frame
is given by the proper time derivatives
of the coordinate time $t$ and radius $r$
\begin{equation}
\label{dttet}
  \partial_t t = \alpha
  \ , \quad
  \partial_t r = \beta
  \ .
\end{equation}
The directed radial derivative
$\partial_r$
is the proper radial derivative
from the point of view of an object
instantaneously at rest in the tetrad frame.
The proper radial derivatives
of the coordinate time $t$ and radius $r$
are
\begin{equation}
\label{drtet}
  \partial_r t = 0
  \ , \quad
  \partial_r r = \gamma
  \ .
\end{equation}
For the more general vierbein of equation~(\ref{genvierbein}),
the directed radial derivative would be
$\partial_r = \nu {\partial / \partial t} + \gamma {\partial / \partial r}$,
and it seen that the gauge choice $\nu = 0$
is equivalent to choosing the time coordinate $t$
so that the proper radial derivative
$\partial_r$
is proportional to the coordinate radial derivative
$\partial / \partial r$.

It is useful to note that
the vierbein coefficients
$(\beta, \gamma)$
constitute the time and radial components
of a covariant tetrad frame 4-vector,
the radial 4-gradient
$\partial_m r$
\begin{equation}
\label{dr}
  \beta
  =
  \partial_t r
  \ , \quad
  \gamma
  =
  \partial_r r
  \ .
\end{equation}
It follows that
$\beta^2 - \gamma^2$
is a Lorentz scalar.
Moreover,
$\beta^2 - \gamma^2$
is gauge-invariant with respect to arbitrary transformations
of the time coordinate $t \rightarrow t(t^\prime, r)$,
the only remaining gauge freedom admitted
by the vierbein of equations~(\ref{vierbein}).
The gauge-invariant scalar
$\beta^2 - \gamma^2$
is related to the mass $M$ interior to $r$, equation~(\ref{M}) below.

The vierbein coefficient $\alpha = \partial_t t$ is not the time component
of a tetrad frame 4-vector because, unlike the radial coordinate $r$,
the time coordinate $t$ changes when the tetrad frame is boosted,
the gauge of time being fixed in accordance with equations~(\ref{gauge}).
In \S\ref{homothetic}
it will be seen that, for self-similar solutions,
the quantity
$t / ( \alpha r ) = \xi^t$
is the time component of the homothetic Killing 4-vector $\xi^m$.

The metric
$\dd s^2 = g_{\mu\nu} \dd x^\mu \dd x^\nu =
\eta_{mn} {e^m}_\mu {e^n}_\nu \dd x^\mu \dd x^\nu$
corresponding to the vierbein of equations~(\ref{vierbein}) is
\begin{equation}
\label{metric}
  \dd s^2
  =
  - \, {\dd t^2 \over \alpha^2}
  + {1 \over \gamma^2} \left( \dd r - {\beta \dd t \over \alpha} \right)^2
  + r^2
  \dd o^2
\end{equation}
where
$\dd o^2 \equiv \dd \theta^2 + \sin^2\!\theta \, \dd \phi^2$
is the metric of the surface of a unit 3-sphere.
It is apparent that the vierbein coefficients $\alpha$ and $\beta$
are related to the lapse and shift in the ADM formalism \cite{ADM62}
(see e.g.\ \cite{Lehner02} for a pedagogical review),
the lapse being $1 / \alpha$, and the shift being $- \beta / \alpha$.
The reciprocal coefficient $1 / \gamma$ might be termed the radial stretch,
since the proper radial interval at fixed time is $\dd r / \gamma$.

An object instantaneously at rest in the tetrad frame satisfies
$\dd r / \dd t = \beta / \alpha$,
according to equation~(\ref{dttet}),
and it then follows from the metric~(\ref{metric})
that the proper time experienced by the object is $\dd t / \alpha$.

In the tetrad formalism,
the covariant derivative
$D_n$
of the tetrad components $p^k$
of a 4-vector is
\begin{equation}
  D_n p^k = \partial_n p^k + \Gamma^k_{mn} p^m
\end{equation}
where the tetrad frame connection coefficients $\Gamma^k_{mn}$,
also known as the Ricci rotation coefficients,
are defined, at least in the case of vanishing torsion, by
the directed derivatives of the tetrad axes
\begin{equation}
\label{gammaGammadef}
  \partial_n \gammavec_m \equiv \Gamma^k_{mn} \gammavec_k
  \ .
\end{equation}
In terms of the vierbein derivatives
$d_{kmn}$
defined by
\begin{equation}
\label{gammad}
  d_{kmn}
  \equiv
  e_{k \kappa} \, {e_n}^\nu \, {\partial {e_m}^\lambda \over \partial x^\nu}
\end{equation}
the connection coefficients
$\Gamma_{kmn} \equiv \eta_{kl} \Gamma^l_{mn}$
with all indices lowered are
\begin{eqnarray}
\label{gammaGamma}
\lefteqn{
  \Gamma_{kmn}
  =
}
&&
\\
\nonumber
&&
  \frac{1}{2}
  \left(
  d_{kmn} - d_{mkn} + d_{nmk} - d_{nkm} + d_{mnk} - d_{knm}
  \right)
  \ .
\end{eqnarray}
The tetrad frame connection coefficients
$\Gamma_{kmn}$
are antisymmetric in their first two indices,
\begin{equation}
\label{Gammaantisymmetric}
  \Gamma_{kmn} = - \Gamma_{mkn}
\end{equation}
which expresses mathematically
the fact that $\Gamma^k_{mn}$ for each given final index $n$
is the generator of a Lorentz transformation
between tetrad frames.
The usual coordinate frame connection coefficients, the Christoffel symbols
$\Gamma_{\kappa\mu\nu} \equiv g_{\kappa\lambda} \Gamma^\lambda_{\mu\nu}$,
are related to the tetrad frame connection coefficients
$\Gamma_{kmn}$ by
\begin{equation}
\label{GammaGamma}
  \Gamma_{\kappa\mu\nu}
  =
  {e^k}_\kappa {e^m}_\mu {e^n}_\nu
  \left(
  \Gamma_{kmn} - d_{kmn}
  \right)
\end{equation}
which has the usual
$\Gamma_{\kappa\mu\nu} = \Gamma_{\kappa\nu\mu}$
symmetry for vanishing torsion.

For the vierbein given by equations~(\ref{vierbein}),
the non-zero tetrad frame connection coefficients are,
in Cartesian coordinates
[coefficients with switched first two indices follow from antisymmetry,
eq.~(\ref{Gammaantisymmetric})],
\begin{subequations}
\label{Gamma}
\begin{eqnarray}
\label{Gammai00}
  \Gamma_{i00}
  &=&
  g \, \hat x_i
\\
\label{Gammai0j}
  \Gamma_{i0j}
  &=&
  h \, \hat x_i \hat x_j
  +
  {\beta \over r} \left( \delta_{ij} - \hat x_i \hat x_j \right)
\\
  \Gamma_{ijk}
  &=&
  {\gamma - 1 \over r}
  \left( \delta_{ik} \hat x_j - \delta_{jk} \hat x_i \right)
\end{eqnarray}
\end{subequations}
where $g$
in equation~(\ref{Gammai00})
is the proper acceleration experienced
by an object that is comoving with the tetrad frame
\begin{equation}
\label{g}
  g \equiv - \, \partial_r \ln\alpha
\end{equation}
and $h$
in equation~(\ref{Gammai0j})
is the proper radial gradient of the proper velocity
between objects each of which is comoving with the tetrad frame at its position
\begin{equation}
\label{h}
  h
  \equiv
  {\partial \beta \over \partial r}
  - \partial_t \ln\gamma
  - \beta \, {\partial \ln\alpha \over \partial r}
  \ .
\end{equation}

The Riemann tensor
$R_{klmn}$
in the tetrad frame
is
defined in the usual way
by the commutator of the covariant derivative,
$\left[ D_k , D_l \right] p_m = R_{klmn} p^n$,
and is given in terms of the tetrad frame connection coefficients by
\begin{eqnarray}
\label{gammaRiemann}
  R_{klmn}
  &=&
    \partial_l \Gamma_{nmk}
  - \partial_k \Gamma_{nml}
  + \Gamma^j_{mk} \Gamma_{njl}
  - \Gamma^j_{ml} \Gamma_{njk}
\nonumber
\\
  &&
  + \, ( \Gamma^j_{lk} - \Gamma^j_{kl} ) \Gamma_{nmj}
\end{eqnarray}
which has two extra terms (the last two) compared to
the usual coordinate expression for the Riemann tensor
in terms of Christoffel symbols.
The Ricci tensor and scalar
are given by the usual contractions
of the Riemann tensor,
$R_{km} = \eta^{ln} R_{klmn}$
and
$R = \eta^{km} R_{km}$,
and the Einstein tensor is given
by the usual expression in terms of the
Ricci tensor and scalar,
$G_{km} = R_{km} - \frac{1}{2} R \, \eta_{km}$.

In the present case
where the vierbein are given by equation~(\ref{vierbein}),
define the interior mass $M$,
the Misner-Sharp \cite{MS64} mass, by
\begin{equation}
\label{M}
  {2 M \over r}
  \equiv
  1 + \beta^2 - \gamma^2
  \ .
\end{equation}
As remarked following equations~(\ref{dr}),
$\beta^2 - \gamma^2$ is a scalar, a gauge-invariant quantity,
and thus so also is the interior mass $M$.
Further, define the symbols $F$, $R$ (not the Ricci scalar!), $P$, and $S$ by
\begin{subequations}
\label{FRPS}
\begin{eqnarray}
\label{F}
  F
  &\equiv&
  {1 \over r} \big( \partial_t \gamma - \beta g \big)
  =
  {1 \over r} \big( \partial_r \beta - \gamma h \big)
  \ ,
\\
  R
  &\equiv&
  {1 \over \gamma} \left(
    {1 \over r^2} \partial_r M
    - \beta F
  \right)
  \ ,
\\
  P
  &\equiv&
  - {1 \over r} \left( \partial_t \beta - \gamma g + {M \over r^2} \right)
  \ ,
\\
  2 S
  &\equiv&
  {r \over \gamma} \biggr[
    \partial_r P + g (R + P)
\nonumber
\\
  &&
    + \, \partial_t F
    + 2 \left( {\beta \over r} + h \right) F
  \biggr]
  \ .
\end{eqnarray}
\end{subequations}
In terms of these quantities,
the components of the Einstein tensor $G_{mn}$ in the tetrad frame are,
in Cartesian coordinates,
\begin{subequations}
\label{einstein}
\begin{eqnarray}
  G_{00}
  &=&
  2 R
  \ ,
\\
\label{einsteini0}
  G_{i0}
  &=&
  - 2 F \hat x_i
  \ ,
\\
  G_{ij}
  &=&
  2 P
  \delta_{ij}
  +
  2 S
  \left( \delta_{ij} - \hat x_i \hat x_j \right)
  \ .
\end{eqnarray}
\end{subequations}
The Einstein tensor embodies the compressive part of the Riemann tensor.
The non-compressive, or tidal, part of the Riemann tensor
is the Weyl tensor $C_{klmn}$, which here simplifies to
\begin{equation}
\label{weyltensor}
  C_{klmn}
  =
  C
  \left[
    \hat t_{kl} \hat t_{mn} - \hat x_{kl} \hat x_{mn}
    + \textstyle\frac{1}{3}
    \left( \eta_{km} \eta_{ln} - \eta_{kn} \eta_{lm} \right)
  \right]
\end{equation}
where $C$ is
\begin{equation}
\label{weylscalar}
  C
  \equiv
  R
  + S
  - {3 M \over r^3}
\end{equation}
and the bivectors (antisymmetric tensors)
$\hat t_{mn}$ and $\hat x_{mn}$
have non-zero components
\begin{equation}
  \hat t_{i0} = - \hat t_{0i} \equiv \hat x_i
  \ , \quad
  \hat x_{ij} \equiv \varepsilon_{ijk} \hat x_k
  \ .
\end{equation}

\subsection{Einstein equations}
\label{einsteinequations}

The Einstein equations in the tetrad frame are
\begin{equation}
\label{einsteineqsraw}
  G_{mn} = 8 \pi T_{mn}
\end{equation}
where $T_{mn}$ is the energy-momentum tensor.
It is apparent from equations~(\ref{einstein})
for the Einstein tensor $G_{mn}$
that the Einstein equations will take their simplest form
when expressed in the center of mass frame,
defined to be the frame in which the momentum density vanishes,
$T_{i0} = 0$.
In the center of mass frame,
$G_{i0}$ vanishes,
and equation~(\ref{einsteini0}) then implies that
the quantity $F$ defined by equation~(\ref{F}) vanishes,
\begin{equation}
  F = 0
  \ ,
\end{equation}
significantly simplifying the expressions for
the other components of the Einstein tensor.
In the center of mass frame,
the most general form of the energy-momentum tensor $T_{mn}$
consistent with spherical symmetry is
\begin{equation}
\label{Tmn}
  T_{00} = \rho
  \ , \quad
  T_{ij} = p_r \, \hat x_i \hat x_j + p_\perp \left( \delta_{ij} - \hat x_i \hat x_j \right)
\end{equation}
where $\rho$ is the proper energy density,
$p_r$ the proper radial pressure,
and $p_\perp$ the proper transverse pressure.
The expressions~(\ref{einstein}) for the Einstein tensor
and (\ref{Tmn}) for the energy-momentum tensor
inserted into the Einstein equations~(\ref{einsteineqsraw})
then imply
\begin{equation}
  R = 4 \pi \rho
  \ , \quad
  P = 4 \pi p_r
  \ , \quad
  S = 4 \pi ( p_\perp - p_r )
  \ .
\end{equation}
The resulting Einstein equations are
\begin{subequations}
\label{einsteineqs}
\begin{eqnarray}
\label{massdensity}
&
\displaystyle{
  {\partial M \over \partial r} = 4 \pi r^2 \rho
  \ ,
}
&
\\
\label{bernoulli}
&
\displaystyle{
  \partial_t \gamma - \beta g = 0
  \ ,
}
&
\\
\label{acceleration}
&
\displaystyle{
  \partial_t \beta - \gamma g + {M \over r^2} + 4 \pi r p_r = 0
  \ ,
}
&
\\
\label{euler}
&
\displaystyle{
  \partial_r p_r - {2 \gamma \over r} ( p_\perp - p_r ) + ( \rho + p_r ) g = 0
  \ .
}
&
\end{eqnarray}
\end{subequations}
Equations~(\ref{einsteineqs}),
coupled with the defining equations~(\ref{g}) for the acceleration $g$
and (\ref{M}) for the interior mass $M$,
constitute the Einstein equations in the most general form
consistent with spherical symmetry.
The first and last equations may be regarded as elliptic constraint equations,
the first equation~(\ref{massdensity})
expressing what appears to be the familiar relation between mass and density,
and the last equation~(\ref{euler})
expressing pressure balance,
the Euler equation.
The midde two equations are hyperbolic evolution equations
governing the evolution of the metric coefficients $\gamma$ and $\beta$.
Equation~(\ref{acceleration})
relates the acceleration $\partial_t \beta$
of the tetrad frame to the gravitational force,
which
comprises the familiar Newtonian force $M / r^2$,
minus a term $\gamma g$ that comes from the acceleration generated
as a result of pressure balance,
plus a term $4 \pi r p_r$ proportional to the radial pressure $p_r$.

It is this last term $4 \pi r p_r$ in the acceleration
equation~(\ref{acceleration})
that makes the interiors of charged black holes problematic and interesting,
for it is this term that,
thanks to the negative radial pressure of the electric field,
tends to make charged black holes gravitationally repulsive
in their deep interiors.

The result of taking $\beta$ times equation~(\ref{acceleration})
minus $\gamma$ times equation~(\ref{bernoulli}) is
\begin{equation}
\label{energyconservation}
  \partial_t M + 4 \pi r^2 p_r \beta = 0
\end{equation}
which expresses energy conservation,
the first law of thermodynamics
(applied to the baryonic fluid, not to the black hole as a whole).
Equations~(\ref{massdensity}) and (\ref{energyconservation})
amply justify the identification of $M$ defined by equation~(\ref{M})
as the mass (energy) interior to radius $r$.

Conservation of energy-momentum
as expressed by the vanishing of the covariant derivative of the
energy-momentum tensor,
\begin{equation}
\label{DT}
  D_m T^{mn} = 0
\end{equation}
is automatically built into Einstein's equations,
a consequence of the Bianchi identities, which ensure that
$D_m G^{mn} = 0$.
The time component, $n = 0$, of equation~(\ref{DT})
gives the energy conservation equation
\begin{equation}
\label{dtrho}
  \partial_t \rho
  + {2 \beta \over r} ( p_\perp - p_r )
  + {( \rho + p_r ) \over r^2} {\partial r^2 \beta \over \partial r}
  = 0
\end{equation}
which can also be derived from Einstein's equations~(\ref{einsteineqs})
by taking the time derivative $\partial_t$ of equation~(\ref{massdensity}),
eliminating $\partial_t M$ using equation~(\ref{energyconservation}),
and then simplifying using the Euler
equation~(\ref{euler}).
The spatial components, $n = 1,2,3$, of equation~(\ref{DT})
give a momentum conservation equation
which reduces precisely to the Euler
equation~(\ref{euler}).

The familiar Oppenheimer-Volkov equation
for general relativistic hydrostatic equilibrium
\begin{equation}
\label{oppenheimervolkov}
  {\partial p \over \partial r} = - {(\rho + p) ( M + 4 \pi r^3 p ) \over r^2 ( 1 - 2 M / r )}
\end{equation}
is recovered from the Einstein equations~(\ref{einsteineqs})
by setting the tetrad frame at rest,
equivalent to setting $\beta \equiv \partial_t r = 0$,
by further assuming isotropic pressure, $p_r = p_\perp$,
and finally by eliminating the acceleration $g$
in the Euler equation~(\ref{euler})
using the acceleration equation~(\ref{acceleration}).

\subsection{Baryonic fluid}
\label{baryons}

As noted by \cite{CT71,OP90,CC99,Gundlach03}
and many others,
the Einstein equations admit similarity solutions
only if pressure and density are proportional.
In this paper we assume that the black hole accretes
a baryonic fluid with an isotropic pressure
and a relativistic equation of state
\begin{equation}
\label{w}
  p_b = w \rho_b
  \ , \quad
  w = \frac{1}{3}
  \ .
\end{equation}
In reality,
the baryonic fluid accreting on to astronomical black holes
is likely to be near but not fully relativistic.
Nevertheless,
given that self-similarity requires $w$ to be constant,
the choice $w = 1/3$ seems the most reasonable.
Moreover,
it may be anticipated that
the fluid will be adiabatically compressed by gravitational
repulsion inside the black hole,
and that it may even undergo a shock if it passes
through an inner horizon.
Both processes will render the baryonic fluid yet more relativistic.

\subsection{Electromagnetic field}
\label{electromageticfield}

The assumption of spherical symmetry
implies that the electromagnetic field can consist only of
a radial electric field $Q / r^2$.
The only non-zero components of
the electromagnetic field tensor $F_{mn}$ in the tetrad frame are then
\begin{equation}
\label{Fi0}
  F_{i0} = - F_{0i} = {Q \over r^2} \hat x_i
  \ .
\end{equation}
Equation~(\ref{Fi0}) may be regarded as defining
what is meant by the charge $Q$ interior to radius $r$.
The electromagnetic energy-momentum tensor in the tetrad frame is given by
$4 \pi T_{mn} = F_{mk} {F_n}^k - {1 \over 4} \eta_{mn} F_{kl} F^{kl}$,
which implies that the density $\rho_e$
and radial and transverse pressures $p_{e,r}$ and $p_{e,\perp}$
of the electric field are
\begin{equation}
\label{rhoe}
  \rho_e = - p_{e,r} = p_{e,\perp} = {Q^2 \over 8 \pi r^4}
  \ .
\end{equation}
Notice that the radial pressure $p_{e,r}$ is negative;
it is this radial tension of the electric field that
leads to gravitational repulsion inside charged black holes.

An electromagnetic field that consists only of a radial electric field
automatically satisfies two of Maxwell's equations,
the source-free ones.
The other two Maxwell's equations, the source ones, are,
in the tetrad frame,
\begin{equation}
\label{maxwell}
  D_n F^{mn}
  =
  \partial_n F^{mn} + \Gamma^m_{nl} F^{ln} + \Gamma^n_{nl} F^{ml}
  =
  4 \pi j^m
  \ .
\end{equation}
For the electromagnetic field given by equation~(\ref{Fi0}),
and with
$j^0 = q$, $j^i = j \hat x_i$,
the Maxwell equations~(\ref{maxwell}) reduce to
\begin{subequations}
\label{dQ}
\begin{eqnarray}
\label{drQ}
  \partial_r Q &=& 4 \pi r^2 q
  \ ,
\\
\label{dtQ}
  \partial_t Q &=& - 4 \pi r^2 j
  \ .
\end{eqnarray}
\end{subequations}
The quantity $q$ is the proper charge density in the fluid frame,
while $j$ is the proper radial current.

There is no need to adjoin the Lorentz force law,
because that is built into Einstein's equations,
which automatically enforce energy-momentum conservation.
Specifically,
if the density and pressure in
the Euler equation~(\ref{euler})
are written as a sum of baryonic and electromagnetic
contributions,
the electromagnetic pressure and density being given
by equations~(\ref{rhoe}),
and if the resulting radial derivative of charge $Q$
is eliminated using the Maxwell equation~(\ref{drQ}),
then the Euler equation~(\ref{euler})
for the charged baryonic fluid becomes
\begin{equation}
  \partial_r p_b - {q Q \over r^2} + (\rho_b + p_b) g = 0
\end{equation}
and it is apparent that the
$q Q / r^2$ term expresses the Lorentz force.

Similarly,
electrical conduction causes Ohmic generation of heat,
but there is no need to doctor the Einstein equations
since they already automatically include this effect.
Specifically,
the electromagnetic contribution to
$\partial_t \rho$
in the energy conservation equation~(\ref{dtrho})
involves a term $\sim (Q \partial_t Q) / r^4 \sim j Q / r^2$
which is precisely the Ohmic dissipation term.

\subsection{Conductivity}

It is still necessary to specify a governing equation for
the radial current $j$.
For diffusive electrical conduction,
the radial current $j$ is proportional to
the radial electric field $Q / r^2$,
with the coefficient of proportionality defining
the electrical conductivity $\sigma$
\begin{equation}
\label{j}
  j = \sigma {Q \over r^2}
\end{equation}
which is just Ohm's law.
A realistic value
(which is not used in this paper)
of the electrical conductivity of a baryonic plasma
at a relativistic temperature $T$ is
\cite{AMY}
\begin{equation}
\label{sigmatrue}
  \sigma =
  {C \over e^2 \ln e^{-1}}
  {k T \over \hbar}
\end{equation}
where $e$ is the dimensionless charge of the electron,
the square root of the fine-structure constant,
and the factor $C \approx 15$ depends on
the mix of particle species.
This electrical conductivity is huge.
A dimensionless measure of the conductivity
(which has units 1/time)
is the conductivity $\sigma$ times the characteristic timescale
$t_\text{BH} \equiv G M / c^3$ of the black hole,
which is of order
\begin{equation}
  \sigma t_\text{BH} \sim
  {T \over T_\text{BH}}
\end{equation}
where $k T_\text{BH} \equiv \hbar / t_\text{BH}$
is the characteristic temperature of the black hole
(for a Schwarzschild black hole, this characteristic temperature
$T_\text{BH}$
is $8 \pi$ times the Hawking temperature).
In the astronomical situation considered here the temperature $T$ of the plasma
is huge compared to the characteristic temperature $T_\text{BH}$ of the black hole.
Indeed if this were not so,
then mass loss by Hawking radiation would tend to compete with mass gain by accretion,
an entirely different situation from the one envisaged here.

In the present paper,
we do not use a realistic value for the conductivity,
equation~(\ref{sigmatrue}),
but instead adopt a phenomenological conductivity
which is greatly reduced compared to any realistic value.
As remarked in the introduction,
the point is that if charge is regarded
as a surrogate for angular momentum,
then electrical conduction can be considered
as an analog of angular momentum transport,
which is intrinsically a much weaker process than electrical conduction.
If it is assumed that the phenomenological conductivity $\sigma$
is some function of the baryonic density $\rho_b$,
then the hypothesis of self-similarity requires that this function
be the square root
(dimensional analysis shows that $\sigma r$ is dimensionless,
and $\rho_b r^2$ is dimensionless)
so
\begin{equation}
\label{sigma}
  \sigma = {\kappa \rho_b^{1/2} \over ( 4\pi )^{1/2}}
\end{equation}
where $\kappa$ is a phenomenological dimensionless coefficient of conductivity,
the factor of $1/(4\pi)^{1/2}$
in equation~(\ref{sigma})
being inserted to simplify the equation when cast in self-similar form,
equation~(\ref{sigmasim}) below.
The realistic conductivity given by equation~(\ref{sigmatrue})
is approximately proportional to
$\rho_b^{1/4}$,
so the notion that the conductivity
increases as some positive power of baryonic density
seems physically reasonable.

It should be noted that the radial current is expected
to increase slightly the radial pressure compared to the transverse pressure
of the charged baryonic fluid,
but this increase is small for diffusive electrical conduction
(and utterly miniscule with the suppressed conductivity adopted here),
and is neglected in this paper.

\section{Similarity solutions}
\label{Similaritysolutions}

\subsection{Similarity hypothesis}

Dimensional analysis
of the spherically symmetric coupled
Einstein-Maxwell equations~(\ref{einsteineqs}) and (\ref{dQ}),
along with the definitions~(\ref{g}) of the acceleration $g$
and (\ref{M}) of the interior mass $M$,
combined with
the equation of state~(\ref{w}) of the baryonic fluid,
the energy-momentum~(\ref{rhoe}) of the electric field,
and the phenomenological conductivity~(\ref{sigma}),
reveals the following quantities to be dimensionless:
\begin{eqnarray}
\label{dimensionless}
&
\displaystyle{
  \eta \equiv {\alpha r \over t}
  \ , \quad
  \beta
  \ , \quad
  \gamma
  \ , \quad
  {M \over r}
  \ , \quad
  {Q \over r}
  \ ,
}
&
\\
\nonumber
&
\displaystyle{
  y \equiv g r
  \ , \quad
  z \equiv 4 \pi r^2 \rho_b
  \ , \quad
  z_q \equiv 4 \pi r^2 q
  \ , \quad
  s \equiv 4 \pi r \sigma
  \ .
}
&
\end{eqnarray}
It is convenient also to denote the dimensionless energy density
$4 \pi r^2 \rho_e$
of the electric field by
\begin{equation}
\label{ze}
  z_e \equiv 4 \pi r^2 \rho_e = {Q^2 \over 2 r^2}
  \ .
\end{equation}
For the phenomenological conductivity $\sigma$ given by equation~(\ref{sigma}),
the dimensionless conductivity $s$ is
\begin{equation}
\label{sigmasim}
  s = \kappa z^{1/2}
  \ .
\end{equation}

If further the gauge of the time coordinate $t$ is fixed in a natural way,
by setting $t$ equal to the proper time experienced by an object
at some boundary,
then the dimension of time $t$ is the same as the dimension of radius $r$,
and then
$\alpha$ and $r/t$ are separately dimensionless,
not just their product $\eta \equiv \alpha r / t$.
In this paper it is not really necessary to choose a specific time
coordinate $t$,
because no observable quantities depend on the choice.
All the same, there is a natural choice available if needed,
which is to synchronize the time to the proper time recorded by
objects (dark matter test particles) that free-fall radially
starting from zero velocity far away.

The dimensionless variables of equation~(\ref{dimensionless})
form a (more than) complete set for the problem at hand,
and it follows
\cite{CC99}
that the spherically symmetric Einstein-Maxwell and subsidiary equations
admit similarity solutions in which the dimensionless variables
are all functions of the dimensionless variable $r/t$.
In particular,
the mass $M$, the charge $Q$, and the radius $r$ of the black hole,
evaluated at some similarity point, such as the outer horizon,
increase linearly with time
\begin{equation}
  M \propto Q \propto r \propto t
  \ .
\end{equation}

Actually the
Einstein-Maxwell and subsidiary equations
would seem to admit more general solutions in which
the dimensionless variables given by equation~(\ref{dimensionless})
are considered to be functions of $r / F(t)$ with $F(t)$
some arbitrary function of time $t$.
However, this apparently greater generality is not genuine,
but merely an artifact of the gauge freedom in the choice
of the time coordinate $t$.
Nevertheless,
the gauge freedom does have the consquence
that the similarity equations
do not involve $\alpha$ and $r/t$ separately,
but only their product, the dimensionless variable
$\eta \equiv \alpha r/t$.

Spacetime points that are fixed in the self-similar frame
have coordinate velocity $\ucoord^r / \ucoord^t = r / t =$ constant,
where $\ucoord^\mu$
is the 4-velocity in the coordinate frame.
In the baryonic tetrad frame these points move with radial velocity
$V = \utet^r / \utet^t$,
where $\utet^m = {e^m}_\mu \ucoord^\mu$
is the 4-velocity in the baryonic tetrad frame
(the 4-velocity $\utet^m$ in the tetrad frames is denoted by a latin $u$,
whereas
the 4-velocity $\ucoord^\mu$ in the coordinate frame
is denoted by a greek upsilon).
With the inverse vierbein ${e^m}_\mu$
given by equations~(\ref{inversevierbein}),
the radial velocity $V$ of the similarity frame
relative to the baryonic tetrad frame is
\begin{equation}
\label{V}
  V
  = {\eta - \beta \over \gamma}
  \ .
\end{equation}
This velocity $V$ is minus the proper velocity
of the baryonic fluid through the self-similar frame.
Horizons occur where the velocity equals the speed of light
\begin{equation}
  V
  =
  \pm 1
  \quad
  \mbox{at horizons}
\end{equation}
while sonic points occur where the velocity equals the sound velocity
\begin{equation}
  V
  =
  \pm \sqrt{w}
  \quad
  \mbox{at sonic points}
  \ .
\end{equation}

\subsection{Homothetic Killing vector}
\label{homothetic}

Similarity solutions are invariant under a scale transformation
of $t$ and $r$ at fixed $r/t$.
The generator of this scale transformation is the homothetic Killing vector
$\xivec$ given by
\cite{CT71,Eardley74,Bogoyavlenskii77,OP90,HP91,Coley97,GNU98,CC99,Harada03}
\begin{equation}
\label{xivec}
  r \xivec \cdot \partialvec
  =
  t {\partial \over \partial t} + r {\partial \over \partial r}
  =
  \left. {\partial \over \partial \ln t} \right|_{r/t}
  =
  r \xi^m \partial_m
  \ .
\end{equation}
The three right hand sides of equation~(\ref{xivec})
express the homothetic vector $\xivec$
in three frames:
the $(t,r)$ coordinate frame;
another coordinate frame in which the time and radial coordinates are
$\ln t$ and $r/t$;
and the baryonic tetrad frame.
The components of the homothetic 4-vector $\xivec$ in the
$(t,r)$
coordinate frame are
$x^\mu / r$,
with time and radial components $( t / r , 1 )$.
Equation~(\ref{xivec})
coupled with the relations~(\ref{directedderivativetr})
between directed and coordinate derivatives imply that
the components $\xi^m$
of the homothetic 4-vector in the baryonic tetrad frame are
\begin{equation}
\label{xi}
  \xi^t = {1 \over \eta}
  \ , \quad
  \xi^r = {V \over \eta}
\end{equation}
where $\eta \equiv \alpha r / t$, and the velocity $V$ is given by
equation~(\ref{V}).
In terms of the homothetic vector,
the relation~(\ref{V}) for the velocity $V$ can be re-expressed as
the statement that the scalar product of the covariant 4-vector $(\beta,\gamma)$
with the contravariant 4-vector $\xi^m$ equals one,
$\beta \xi^t + \gamma \xi^r = 1$.
The magnitude squared of the homothetic vector
defines the homothetic scalar $H$
\begin{equation}
\label{H}
  H \equiv
  - \xi_m \xi^m
  =
  {1 - V^2 \over \eta^2}
\end{equation}
which, like the interior mass $M$,
is gauge-invariant.
Horizons occur where the homothetic scalar vanishes, $H = 0$,
which demonstrates that the location of horizons is independent of the
choice of the coordinate system or of the frame of reference,
which is as it should be.

Associated with the scale transformation symmetry is a conservation law.
To derive this conservation law,
consider the Lagrangian for a (neutral) test particle
freely-falling with coordinate 4-velocity
$\ucoord^\mu \equiv \dd x^\mu / \dd \tau$
\begin{equation}
  {\cal L} = g_{\mu\nu} \ucoord^\mu \ucoord^\nu
  \ .
\end{equation}
If the metric~(\ref{metric}) is cast in terms of coordinates
$\ln t$ and $r / t$
in place of $t$ and $r$,
then all of the metric coefficients become proportional to $t^2$
at fixed $r / t$, so that
$\partial {\cal L} / \partial \ln t = 2 {\cal L} = - 2$,
the last equality being true at least for a massive particle,
for which $\ucoord_\mu \ucoord^\mu = -1$.
The equation of motion for the coordinate $\ln t$
of the freely-falling particle is
\begin{equation}
  {\dd \over \dd \tau} {\partial {\cal L} \over \partial \ucoord^{\lnt}}
  -
  {\partial {\cal L} \over \partial \ln t}
  = 0
\end{equation}
which then reduces to
\begin{equation}
\label{dudlnt}
  {\dd \ucoord_{\lnt} \over \dd \tau} + 1 = 0
  \ .
\end{equation}
Equation~(\ref{dudlnt}) integrates to
\begin{equation}
\label{ulntt}
  \ucoord_{\lnt} + \tau = 0
\end{equation}
where the constant of integration has been absorbed
into a shift of the zero point of the proper time $\tau$.
Equation~(\ref{ulntt}) is the sought-for conservation law:
it says that the homothetic momentum
$\ucoord_{\lnt}$
associated with the coordinate $\lnt$
of a freely-falling particle
is equal to minus the proper time $\tau$ along the geodesic,
provided that the zero point of proper time is set appropriately.
In the tetrad frame, the homothetic momentum
$\ucoord_{\lnt}$
is
\begin{equation}
\label{ulnt}
  \ucoord_{\lnt} = r \, \xi_m \utet^m
  \ .
\end{equation}
Although the above derivation was for a massive particle,
the results~(\ref{ulntt}) and (\ref{ulnt}) remain valid also for
a massless particle,
with the understanding that for a massless particle
the proper time $\tau$ is a constant along the null geodesic.

\subsection{Ingoing vs.\ Outgoing}
\label{inoutgoing}

In the Reissner-Nordstr\"om (RN) geometry,
geodesics between the outer and inner horizons
can be classified as ingoing or outgoing according
to the sign of their specific energy
$E = - \ucoord_t$,
where $\ucoord^\mu$ is the coordinate 4-velocity along the geodesic.
Outgoing geodesics have negative energy,
and go backwards in RN time $t$.
The horizons of the RN geometry
can likewise be classified as ingoing or outgoing.
Only ingoing geodesics (those with $E > 0$) can pass through ingoing horizons,
while only outgoing geodesics (those with $E < 0$) can pass through outgoing
horizons.
Timelike or null geodesics
that start from outside the outer horizon
necessarily go forward in time and have positive energy,
and are therefore necessarily ingoing,
and cannot pass through the outgoing inner horizon,
the so-called Cauchy horizon.
Charged particles do not follow geodesics,
being accelerated by the electric field,
and particles with the same sign of charge as the black hole,
and with a large enough charge-to-mass ratio,
do routinely switch from ingoing to outgoing
between the outer and inner horizons,
and thus do fall through the outgoing Cauchy horizon.

In the similarity solutions,
geodesics between the outer and inner horizons
can be classified as ingoing or outgoing according
to the sign of minus the homothetic momentum $- \ucoord_{\lnt}$,
equation~(\ref{ulnt}).
This definition of ingoing versus outgoing is gauge-invariant,
independent of the choice of either coordinates or tetrad frame.
Note that according to equation~(\ref{ulntt})
$\ucoord_{\lnt}$ can only decrease along a geodesic,
since the proper time $\tau$ can only increase.
Thus a freely-falling (neutral) massive particle can change from outgoing
($\ucoord_{\lnt} > 0$)
to ingoing
($\ucoord_{\lnt} < 0$),
but once it is ingoing it must remain ingoing thereafter,
as long as it continues in free-fall.
The asymmetry between ingoing and outgoing
reflects the fact that the similarity solution
is not invariant with respect to time reversal.

The tetrad frame can be classified as ingoing or outgoing according to
whether objects instantaneously at rest in that frame are ingoing or outgoing.
The 4-velocity of an object instantaneously at rest in the tetrad frame is
$\utet^t = 1$ and $\utet^r = 0$,
and in this case equation~(\ref{ulnt}) simplifies to
$\ucoord_{\lnt} = - {r / \eta}$.
Thus the tetrad frame
is ingoing or outgoing according to the sign of
$\eta \equiv \alpha r / t$.
When the tetrad frame switches between ingoing and outgoing,
so that $\eta$ changes sign,
the time component $\xi^t = 1/\eta$ of the homothetic vector
must pass through zero;
it cannot pass through infinity
because the homothetic vector $\xi^m$ is finite.
At the same time, the radial component $\xi^r = V/\eta$
must remain finite, neither zero nor infinite,
again because the homothetic vector $\xi^m$ is finite,
neither identically zero nor infinite.
It follows that when the tetrad frame switches between ingoing and outgoing,
$\eta$ must pass through $\pm \infty$,
and the velocity $V$ must simultaneously pass through $\pm \infty$,
changing sign simultaneously with $\eta$.
Consistency of the self-similar solutions requires that $\xi^r$
always remain of the same sign,
since if $\xi^r$ changed sign it would indicate that the fluid turns
back on itself, which physically cannot happen
(at least for baryons).
In the black hole solutions of the present paper,
$\xi^r$ is initially positive ($\eta$ and $V$ are both positive).
Thus one concludes that
the baryonic tetrad frame is ingoing if $V$ is positive,
and outgoing if $V$ is negative,
\begin{equation}
  \begin{array}{cl}
    V > 0 & \qquad \mbox{ingoing} \\
    V < 0 & \qquad \mbox{outgoing}
  \end{array}
\end{equation}
and that the passage between ingoing and outgoing is marked
by $V$ passing through $\pm \infty$.
It is this condition on the velocity $V$ that will be used in the
results section~\ref{Resultsbaryons}
to indicate whether the baryonic fluid is ingoing or outgoing.

There is no physical divergence or discontinuity in the properties
of the baryonic fluid when
the velocity $V$ passes through $\pm \infty$.
All that happens is that the homothetic 4-vector $\xivec$
in the baryonic tetrad frame
switches between pointing forwards in time (ingoing, positive $\xi^t$)
to backwards in time (outgoing, negative $\xi^t$).

\subsection{Geodesics}
\label{geodesics}

Towards the end of this paper,
\S\ref{Appearance},
the visceral question is posed:
What does it actually look like if you fall inside
one of the black holes described in this paper?
To address this question requires ray-tracing,
which requires solving for geodesics,
notably null geodesics.

The equations of motion for freely falling test particles,
massive or massless,
take their most transparent form when expressed in self-similar,
or homothetic, coordinates,
where the homothetic symmetry is explicit.
Let $t_h$ denote the homothetic time,
the time coordinate defined by the gauge choice~(\ref{gauge})
when the tetrad frame is chosen to be comoving,
possibly superluminally, with the similarity frame.
Define the self-similar coordinate $X$ by
\begin{equation}
\label{dX}
  \dd X
  \equiv
  {H \, \dd \ln ( r / t_h ) \over \gamma \xi^t + \beta \xi^r}
\end{equation}
where $H$ is the homothetic scalar,
given by equation~(\ref{H}).
The factor of $H / ( \gamma \xi^t + \beta \xi^r )$ is included
in equation~(\ref{dX})
so that $\ucoord^X \equiv \dd X / \dd \tau$,
see equation~(\ref{uX}) below,
remains finite and well-behaved everywhere along a geodesic.
Note that $\gamma \xi^t + \beta \xi^r$ is a scalar,
the scalar product of the covariant 4-vector $(\gamma, \beta)$
with the contravariant 4-vector $(\xi^t, \xi^r)$.
The metric with respect to homothetic coordinates
$\ln t_h$
and
$X$
is
\begin{equation}
\label{metrich}
  \dd s^2
  =
  r^2 \left(
  - H \, \dd \ln t_h^2
  + {\dd X^2 \over H}
  + \dd o^2
  \right)
  \ .
\end{equation}

Without loss of generality,
let a test particle move in the $\theta = \pi/2$ plane,
with varying azimuthal angle $\phi$.
In homothetic coordinates
$\ln t_h$
and
$X$,
the 4-velocity $\ucoord^\mu$
satisfies
\begin{subequations}
\label{uhomothetic}
\begin{eqnarray}
\label{uhomotheticlnt}
  \ucoord_{\lnt_h} &=& - \, \tau
\\
\label{uhomotheticphi}
  \ucoord_\phi &=& J
\\
\label{uhomotheticuu}
  \ucoord_\mu \ucoord^\mu &=& - \mu^2
  \ .
\end{eqnarray}
\end{subequations}
The first of equations~(\ref{uhomothetic})
is equation~(\ref{ulntt}) for the homothetic momentum,
as derived in \S\ref{homothetic}.
The second equation expresses
conservation of angular momentum per unit energy $J$.
The last equation expresses conservation of rest mass per unit energy $\mu$,
which can be taken to be $\mu = 1$ for a massive particle,
or $\mu = 0$ for a massless particle.

From equations~(\ref{uhomothetic}) and the metric~(\ref{metrich})
it follows that in homothetic coordinates
the coordinate 4-velocity $\ucoord^\mu$
of a particle with proper time $\tau$, angular momentum per unit energy $J$,
and mass per unit energy $\mu$ is given by
\begin{subequations}
\label{momentumh}
\begin{eqnarray}
\label{uth}
  \ucoord^{\lnt_h}
  &=&
  {\tau \over H r^2}
\\
\label{uX}
  \ucoord^X
  &=&
  \pm
  {1 \over r^2}
  \left[
    \tau^2 - H ( J^2 + \mu^2 r^2 )
  \right]^{1/2}
\\
\label{uphi}
  \ucoord^\phi
  &=&
  {J \over r^2}
\end{eqnarray}
\end{subequations}
in which the sign of
$\ucoord^X$
is determined according to whether the particle is moving outwards
or inwards relative to the similarity frame.
Relative to an observer at rest in the tetrad frame,
the tetrad 4-velocity $\utet^m$ of the particle is
\begin{subequations}
\label{ugeo}
\begin{eqnarray}
  \utet^t
  &=&
  {\xi^t \tau \pm \xi^r \left[ \tau^2 - H ( J^2 + \mu^2 r^2 ) \right]^{1/2}
  \over H r}
\\
\label{ugeor}
  \utet^r
  &=&
  {\xi^r \tau \pm \xi^t \left[ \tau^2 - H ( J^2 + \mu^2 r^2 ) \right]^{1/2}
  \over H r}
\\
  \utet^\perp
  &=&
  {J \over r}
\end{eqnarray}
\end{subequations}
where $\xi^m$ are the components of the homothetic 4-vector in the tetrad frame,
equation~(\ref{xi}),
and the $\pm$ signs in the expressions for $\utet^t$ and $\utet^r$
are the same as the sign of
$\ucoord^X$, equation~(\ref{uX}).

The shape of a geodesic trajectory can be obtained by integrating
$\dd \phi / \dd X =  \ucoord^\phi / \ucoord^X$.
For photons, the proper time $\tau$ is constant,
and the rest mass $\mu$ is zero,
and the equation for the angle $\phi$ along a null geodesic
reduces to an integral
\begin{equation}
\label{phi}
  \phi
  =
  \int
  {J \, \dd X \over
  \left( \tau^2 - H J^2 \right)^{1/2}}
  \ .
\end{equation}
A null geodesic passes through
peri- or apo-apsis
in the self-similar frame
where the denominator of the integrand of equation~(\ref{phi}) vanishes.
The separatrix between null geodesics
that do or do not fall into the black hole,
the equivalent of the photon sphere,
occurs where the denominator not only vanishes, but is a double root,
which happens when
(the subscript ph signifies the photon sphere equivalent)
\begin{equation}
\label{Hph}
  H_\textrm{ph} = {\tau_\textrm{ph}^2 \over J_\textrm{ph}^2}
  \ , \quad
  \left. {\dd H \over \dd X} \right|_{H = H_\textrm{ph}} = 0
  \ .
\end{equation}

An observer sees a particle, either a photon or a massive particle,
to come from angle $\chi$ away from the direction
to the center of the black hole.
The angle
$\chi \in [0,\pi]$
is given by
\begin{equation}
  \tan \chi =
  {\utet^\perp \over \utet^r}
  =
  {J \left\{ \xi^r \tau
    \mp \xi^t \left[ \tau^2 - H ( J^2  + \mu^2 r^2 ) \right]^{1/2} \right\}
  \over
  (\xi^t)^2 ( J^2 + \mu^2 r^2 ) - \tau^2}
\end{equation}
in which the $\mp$ sign is opposite to the $\pm$ sign
in the expression~(\ref{ugeor}) for $\utet^r$.

The apparent edge of the black hole on the sky is set by
photons from the photon sphere equivalent,
whose proper time
$\tau_\textrm{ph}$
and angular momentum
$J_\textrm{ph}$
satisfy equations~(\ref{Hph}).
The angular size
$\chi_\textrm{ph} \in [0,\pi]$
of the black hole on the sky is then
\begin{equation}
\label{chiph}
  \tan \chi_\textrm{ph} =
  {\xi^r H_\textrm{ph}^{1/2}
  \mp
  \xi^t \left( H_\textrm{ph} - H \right)^{1/2}
  \over
  (\xi^t)^2 - H_\textrm{ph}}
\end{equation}
in which the $\mp$ sign is negative for observers
outside the radius of the photon sphere equivalent,
$r > r_\textrm{ph}$
(the photons move outward in the similarity frame),
and positive for observers inside,
$r < r_\textrm{ph}$
(the photons move inward in the similarity frame).

The observed blueshift of photons at the edge of the black hole equals
the ratio $\utet_\textrm{ph}^t / \utet_\textrm{ph,0}^t$
of the observed-to-emitted energy of a photon from the photon sphere equivalent.
We take the emitted energy $\utet_\textrm{ph,0}^t$
to be the energy of the photon from the point of view of an observer
at rest in the self-similar frame at the radius
of the photon sphere equivalent,
$r_\textrm{ph}$.
The observed blueshift factor is then
\begin{equation}
\label{blueph}
  {\utet_\textrm{ph}^t \over \utet_\textrm{ph,0}^t}
  =
  {r_\textrm{ph} \left[ H_\textrm{ph} + ( \xi^r )^2 \right]
  \over
  r \left[
  \xi^t H_\textrm{ph}^{1/2}
  \mp
  \xi^r \left( H_\textrm{ph} - H \right)^{1/2}
  \right]}
\end{equation}
where again the $\mp$ sign is negative for observers
outside the radius of the photon sphere equivalent,
$r > r_\textrm{ph}$,
and positive for observers inside,
$r < r_\textrm{ph}$.

In the limit of small accretion rates and small conductivities
(and in the absence of mass inflation),
the self-similar geometry asymptotes to
the Reissner-Nordstr\"om geometry
over any limited range of time,
and $X$ and $H$ tend to
\begin{equation}
\label{XHlimit}
  X
  \rightarrow
  - \,
  {\tau \over r}
\end{equation}
\begin{equation}
  H
  \rightarrow
  {\tau^2 \over r^2}
  \left( 1 - {2 M \over r} \right)
\end{equation}
where $1 - 2 M / r = 1 - 2 \Mc / r + Q^2 / r^2$
with $\Mc$ and $Q$ respectively the constant mass and charge
of the RN black hole.
In this limit,
$H(X)$ is a cubic polynomial in $X$ for an uncharged black hole, $Q = 0$,
or a quartic polynomial in $X$ for a charged black hole,
and the ray-tracing integral~(\ref{phi}) becomes an elliptic integral,
just as in the Schwarzschild and Reissner-Nordstr\"om geometries.

\subsection{Shock jump conditions}
\label{shock}

The baryonic fluid may undergo a relativistic shock
\cite{CT71,Bogoyavlenskii77},
where the density, pressure, and velocity of the baryons
change discontinuously across a shock front,
a 3-dimensional hypersurface in spacetime.
Let $n^m$ denote the (spacelike) normal to the shock front.
Conservation of energy-momentum imposes the shock jump conditions
\begin{equation}
\label{jump}
  \left[ n_m T^{mn} \right]_-^+ = 0
\end{equation}
where $[\,]_-^+$ denotes the difference between
post-shock ($+$)
and pre-shock ($-$) values.
For a spherical shock wave,
the shock normal $n^m$ in the rest frame of the shock front has components
$n^t = 0$, $n^r = 1$.
The energy-momentum tensor $T_e^{mn}$ of the radial electric field
remains invariant under any radial Lorentz boost,
so for a spherical shock wave the shock jump conditions reduce
to jump conditions on the baryonic energy-momentum tensor $T_b^{mn}$ alone
\begin{equation}
\label{jumpb}
  \left[ n_m T_b^{mn} \right]_-^+ = 0
  \ .
\end{equation}

Let $V_-$ and $V_+$ denote
the proper velocities of the shock front relative respectively to the
pre-shock ($-$) and post-shock ($+$) baryons.
If the shock is self-similar,
as considered in the present paper,
then the velocities $V_{\pm}$ are equal to the velocities
of the self-similar frame relative to the baryons,
but equations~(\ref{vshk})--(\ref{gpbshk}) below
remain valid also for a general spherical shock.
The shock jump conditions~(\ref{jumpb}) imply that
the pre- and post-shock velocities $V_{\pm}$ are related by
\begin{equation}
\label{vshk}
  V_- V_+ = w
\end{equation}
and that the ratio
of post- to pre-shock densities is
\begin{equation}
  {\rho_{b,+} \over \rho_{b,-}}
  =
  {V_- ( 1 - V_+^2 ) \over V_+ ( 1 - V_-^2 )}
  \ .
\end{equation}
Relative to the pre-shock baryons, the post-shock baryons are
Lorentz boosted by 4-velocity $\utet^m$ given by
\begin{eqnarray}
\label{ushk}
  \utet^t - \utet^r
  &=&
  \left[ {( 1 - V_- ) ( 1 + V_+ ) \over ( 1 + V_- ) ( 1 - V_+ )} \right]^{1/2}
  \ ,
\nonumber
\\
  \utet^t + \utet^r
  &=&
  {1 \over \utet^t - \utet^r}
  \ .
\end{eqnarray}
The vierbein coefficients $\beta$ and $\gamma$,
which form the time and space components of a covariant 4-vector,
equation~(\ref{dr}),
are Lorentz boosted by the 4-velocity $\utet^m$ across the shock.
Written in a form that remains numerically well-behaved
even under extreme conditions,
the relation between post- and pre-shock coefficients is
\begin{equation}
\label{gpbshk}
  \gamma_+ \pm \beta_+
  =
  ( \gamma_- \pm \beta_- )
  ( \utet^t \pm \utet^r )
  \ .
\end{equation}
The homothetic Killing vector $\xi^m$, a contraviant 4-vector,
is similarly Lorentz boosted across the shock:
\begin{equation}
  \xi^t_+ \pm \xi^r_+
  =
  ( \xi^t_- \pm \xi^r_- )
  ( \utet^t \mp \utet^r )
  \ .
\end{equation}
The scalars
$\beta^2 - \gamma^2$
and
$H = - \xi_m \xi^m$
are unchanged by a Lorentz boost,
and are therefore continuous across the shock.
The interior mass $M$,
which is related to the scalar $\beta^2 - \gamma^2$ by equation~(\ref{M}),
is consequently also continuous across the shock,
as one might have expected.
The interior charge $Q$ is continuous across the shock,
because the energy-momentum tensor of the electric field,
equation~(\ref{rhoe}), is invariant under a radial Lorentz boost.

\subsection{Integrals of the similarity equations}

The similarity hypothesis requires that
all dimensionless quantities must be some function of
a single dimensionless variable,
which can be taken to be for example $r/t$.
The directed time and radial derivatives
of a dimensionless quantity $f(r/t)$ are
\begin{equation}
\label{dselfsim}
  r \partial_t f = ( - \eta + \beta ) {\dd f \over \dd \ln(r/t)}
  \ , \quad
  r \partial_r f = \gamma {\dd f \over \dd \ln(r/t)}
  \ .
\end{equation}
In particular,
any dimensionless quantity $f(r/t)$ must satisfy
\begin{equation}
\label{fsim}
  \left( \partial_t + V \partial_r \right) f
  =
  0
\end{equation}
where $V$, given by equation~(\ref{V}),
is the velocity of the self-similar frame
relative to the tetrad frame.
In terms of the homothetic Killing vector $\xi^m$ given by equation~(\ref{xi}),
equation~(\ref{fsim}) becomes
\begin{equation}
\label{xifsim}
  \xi^m \partial_m f
  =
  0
  \ .
\end{equation}

The ordinary differential equations
determining the self-similar evolution of the baryonic fluid
admit three integrals.
The first integral follows from
\begin{equation}
\label{Msim}
  \xi^m \partial_m (M / r) = 0
\end{equation}
as a particular case of equation~(\ref{xifsim}).
Inserting equations~(\ref{massdensity}) and (\ref{energyconservation})
into equation~(\ref{Msim}),
and simplifying,
yields an equation for the dimensionless ratio $M / r$
of interior mass to radius
\begin{equation}
\label{Mr}
  {M \over r}
  =
  z ( \gamma \xi^r - w \beta \xi^t )
  +
  z_e
  \ .
\end{equation}
Equation~(\ref{Mr})
expresses $M/r$,
defined in terms of $\beta$ and $\gamma$ by equation~(\ref{M}),
in terms of other dimensionsionless variables.
In the present paper
we do not impose equation~(\ref{Mr}) as one of the evolution equations,
but rather use it as a check on the accuracy of the numerical integration.
Usually the check is satisfied to $\sim 10^{-9}$ or better,
but in some cases the check fails badly,
namely in some cases where the integration terminates
at an irregular sonic point below the Cauchy horizon.
Where the check fails,
the problem is that
$\gamma \xi^r - w \beta \xi^t$
on the right hand side of equation~(\ref{Mr})
is a tiny difference of two large numbers, which leads to loss of precision.
However,
the problem is clearly with the check equation~(\ref{Mr}),
not with the equations being integrated,
which is the reason for not including equation~(\ref{Mr})
among the evolution equations.

The second integral of the similarity equations follows from
\begin{equation}
\label{Qsim}
  \xi^m \partial_m (Q / r) = 0
\end{equation}
which,
when the Maxwell's equations~(\ref{dQ}) are inserted,
yields an equation for the dimensionless charge density
$z_q \equiv 4 \pi r^2 q$
\begin{equation}
\label{zq}
  z_q
  =
  {Q ( 1 + s \xi^t ) \over r \xi^r}
  \ .
\end{equation}

The third integral of the similarity equations follows from
\begin{equation}
\label{psim}
  w \, 4 \pi r^2 (\rho + p_r) \xi^m \partial_m \beta
  +
  \gamma V  \xi^m \partial_m ( 4 \pi r^2 p_r )
  =
  0
\end{equation}
as another consequence of equation~(\ref{xifsim}).
Equation~(\ref{psim}) simplifies as follows.
First recast the $\partial_t p_r$ part of
$\xi^m \partial_m p_r$
in terms of $\partial_t \rho_b$ and $\partial_t Q$
using the baryonic and electromagnetic
equations of state~(\ref{w}) and (\ref{rhoe}).
Then eliminate the derivatives
$\partial_t \beta$,
$\partial_r p_r$,
$\partial_t \rho$,
and
$\partial_t Q$
using equations~(\ref{acceleration}),
(\ref{euler}),
(\ref{dtrho}),
and
(\ref{dtQ}),
along with the conductivity equation~(\ref{j}).
The result is an expression that contains no derivatives.
Translating into the dimensionless variables of
equations~(\ref{dimensionless}) and (\ref{ze})
yields an equation for the dimensionless proper acceleration
$y \equiv g r$
\begin{equation}
\label{y}
  y
  =
  {\xi^r \left\{
  \begin{array}{l}
    2 w M / r
    + 2 z_e \left[ (1 {-} w) + (1 {+} w) s \xi^t \right]
    \\
    - \, w \left[ (1 {+} w) z \xi^t \right]^2
  \end{array}
  \right\}
  \over
  (1 {+} w) z \left[ (\xi^r)^2 - w (\xi^t)^2 \right]}
  \ .
\end{equation}
The denominator of expression~(\ref{y})
for the dimensionless acceleration $y$
is proportional to
$V^2 - w$,
which is zero when the baryonic fluid velocity through the similarity frame
equals the speed of sound,
$V = \pm \sqrt{w}$.
Generally, there are two possibilities for
what happens when the fluid velocity passes the sound barrier,
depending on whether the velocity accelerates or decelerates.
If the fluid velocity accelerates from subsonic to supersonic,
then information can propagate upstream from the sonic point,
damping any tendency to develop large accelerations,
and allowing the fluid to pass smoothly through the sonic point
where $V = \pm \sqrt{w}$.
If on the other hand the fluid velocity decelerates from supersonic to subsonic,
then information cannot propagate upstream,
and in general the fluid cannot pass smoothly through a sonic point.
Instead, the fluid steepens into a shock,
and the velocity $V$ changes discontinuously
from being supersonic, $|V| > \sqrt{w}$, to being subsonic, $|V| < \sqrt{w}$.

\subsection{Similarity differential equations}

Although $r / t$ would seem to be a natural choice of
dimensionless integration variable,
in practice it proves to be a poor choice,
because $r / t$ does not vary monotonically,
but rather oscillates through zero
(the time coordinate $t$ oscillates through $\pm \infty$),
sometimes many times,
as the baryonic fluid inside the black hole
transitions between ingoing and outgoing.
A suitable alternative choice of dimensionless integration variable is
the dimensionless time parameter $x$ defined by
\begin{equation}
  \dd x \equiv {\dd \tau \over r}
\end{equation}
evaluated along the path of the baryonic fluid.
The dimensionless time parameter $x$ naturally increases monotonically,
since the proper time $\tau$ does.
The baryonic proper time $\tau$,
the time coordinate $t$,
and the radial coordinate $r$
evolve along the path of the baryonic fluid as
\begin{subequations}
\begin{eqnarray}
\label{taux}
  {\dd \tau \over \dd x}
  &=&
  r
  \ ,
\\
\label{tx}
  {\dd \ln t \over \dd x}
  &=&
  \eta
  \ ,
\\
\label{rx}
  {\dd \ln r \over \dd x}
  &=&
  \beta
  \ .
\end{eqnarray}
\end{subequations}
The time coordinate $t$ is not used in
the results section, \S\ref{Resultsbaryons},
but the differential equation~(\ref{tx}) governing its evolution
is given for completeness.
Equation~(\ref{tx}) presumes that the gauge of baryonic time $t$ is
chosen in the natural way,
such that the units of time are the same as the units of radius,
so that $r / t$ is a dimensionless variable.

An overcomplete set of equations
[only three of the four equations~(\ref{dsim}) below are independent,
the four variables $\xi^t$, $\xi^r$, $\gamma$, and $\beta$
being related by
$\beta \xi^t + \gamma \xi^r = 1$,
equation~(\ref{xib}) below]
governing the self-similar evolution of the remaining variables is
\begin{subequations}
\label{dsim}
\begin{eqnarray}
\label{xitsim}
  {\dd \xi^t \over \dd x}
  &=&
  - \, y \xi^r + \gamma \xi^r
\\
\label{xirsim}
  {\dd \xi^r \over \dd x}
  &=&
  - \, y \xi^ t
  - \beta \xi^r
  + (1 {+} w) z \xi^t \xi^r
\\
  {\dd \gamma \over \dd x}
  &=&
  \beta y
\\
  {\dd \beta \over \dd x}
  &=&
  \gamma y
  - (1 {+} w) z \gamma \xi^r
\end{eqnarray}
\end{subequations}
together with
\begin{subequations}
\label{dsimz}
\begin{eqnarray}
  {\dd \ln [ r^{1 + 3 w} z ( \xi^r )^{1 + w} ] \over \dd x}
  &=&
  {2 z_e s \over z}
  \qquad
\\
  {\dd \ln Q \over \dd x}
  &=&
  - \, s
  \ .
\end{eqnarray}
\end{subequations}
To maintain numerical precision,
it is important to avoid expressing small quantities
as differences of large quantities.
A suitable choice of variables to integrate
is
$\xi^t + \xi^r$,
$\beta - \gamma$,
$\gamma$,
each of which can be tiny in some circumstances.
Starting from these variables,
the following chain of equations yield
the remaining variables in a fashion that ensures numerical precision:
\begin{subequations}
\begin{eqnarray}
\label{xib}
&
\displaystyle{
  \xi^t - \xi^r
  =
  {2 - ( \beta + \gamma ) ( \xi^t + \xi^r ) \over \beta - \gamma}
}
&
\\
&
\displaystyle{
  H
  =
  ( \xi^t + \xi^r ) ( \xi^t - \xi^r )
}
&
\\
&
\displaystyle{
  {2 M \over r}
  =
  1 + ( \beta + \gamma ) ( \beta - \gamma )
  \ .
}
&
\end{eqnarray}
\end{subequations}

For reference,
the differential equation for $X$,
useful in ray-tracing, see equation~(\ref{phi}), is
\begin{equation}
  {\dd X \over \dd x} = - \, \xi^r
  \ .
\end{equation}

\subsection{Boundary conditions at the outer sonic point}
\label{boundaryconditions}

The boundary conditions of the calculation are set in this paper
at an outer boundary,
taken to be a regular sonic point,
outside the outer horizon of the black hole,
where the infalling baryonic fluid transitions smoothly
from subsonic to supersonic velocity.
The behaviour in the vicinity of sonic points
in general relativistic similarity solutions has been discussed by
\cite{Bogoyavlenskii77,BH78,OP90,CY90,CC00,CCGNU01,Harada01}.

In setting the boundary conditions at the outer sonic point,
we choose not to enquire how the accreting fluid managed to get to that point,
but simply assume that astrophysical processes can arrange themselves
so as to feed the black hole in a steady self-similar fashion.
Light curves of real accreting black holes are typically variable
rather than steady
\cite{UM04,ZM04},
which could indicate that accretion flows are typically non-steady,
but we ignore this difficulty,
assuming that at least in principle a black hole could accrete steadily.
Integrating outwards from the sonic point
reveals that the similarity solutions typically do not extend to infinity,
but rather terminate,
either at a stagnation point where the velocity $V$ is zero
and the fluid wants to turn back on itself,
or at an irregular sonic point where the acceleration diverges.
Thus the similarity solutions considered in this paper
are typically not complete self-consistent solutions
valid to arbitrary distances from the black hole.
Again, we choose to ignore this difficulty,
noting that the conditions for self-similarity,
such as a relativistic equation of state,
could well break down,
so the failure of the solutions to extend to infinity
is not necessarily a fatal difficulty.

At the sonic point,
where the fluid velocity equals the sound speed,
$V = \sqrt{w}$,
the denominator of the expression~(\ref{y})
for the dimensionless acceleration $y$ is zero,
and the numerator must simultaneously vanish for the acceleration
to remain finite.
The vanishing of the numerator and denominator of
the right hand side of equation~(\ref{y})
imposes two boundary conditions at the sonic point,
and a third condition follows if $y$ is taken to be not only
continuous but also differentiable at the sonic point.
Physically,
sound waves generated by discontinuities near the sonic point
can propagate upstream, modifying the flow so as
to ensure a smooth transition through the sonic point.
The value of the dimensionless acceleration $y$ at the sonic point,
where the numerator and denominator of equation~(\ref{y}) vanish,
is given by the ratio of the derivatives of the numerator and denominator,
according to L'H\^opital's rule.
If these derivatives are expanded
according to equations~(\ref{dsim}) and (\ref{dsimz}),
then the result is an equation of the form
\begin{equation}
\label{ysonic}
  y = {a + b y \over c + d y}
\end{equation}
where $a$, $b$, $c$, and $d$ are functions of dimensionless variables
not including $y$.
Equation~(\ref{ysonic}) is a quadratic equation for $y$,
and generically there are two solutions,
if solutions exist.
Numerically,
when two solutions exist,
one represents a transition from subsonic to supersonic,
while the other represents a transition from supersonic to subsonic;
it is the former that is physically relevant.

For a sonic point at the outer boundary to be acceptable,
it must satisfy four conditions.
First, the sonic point must be regular,
which requires that a solution to equation~(\ref{ysonic}) exist.
Second, the baryons must fall inward, that is,
the velocity $V$ of the similarity frame relative to the baryons
must be positive, $V = + \sqrt{w}$.
Third,
the velocity $V$ must transition from subsonic to supersonic
as the baryons fall inward,
which is a condition on the derivative of $V$.
Fourth,
the radial 4-gradient
$(\beta, \gamma) \equiv (\partial_t r, \partial_r r)$
must be spacelike,
$\gamma^2 - \beta^2 > 0$,
which, as proven in Appendix~\ref{proof},
is a necessary condition that the boundary be causally connected to
(not separated by a horizon from)
a hypothetical asymptotically flat empty region of space at large radius $r$.
In the cases we have investigated,
it is the fourth condition,
$\gamma^2 - \beta^2 > 0$,
that sets limits on the values of physical parameters
at the outer sonic point.

Given the assumption that the outer boundary is a regular sonic point,
there are two further boundary conditions to be set for the baryons:
the dimensionless accretion rate $\eta_s \equiv \alpha r / t$,
and the charge-to-mass ratio $Q/\Mc$ of the black hole,
where $\Mc$ is the charge-augmented mass defined by equation~(\ref{Mc}) below.

The dimensionless accretion rate $\eta_s = \alpha r / t$ at the sonic point
is roughly speaking the velocity at which the black hole is expanding,
the coefficient $\alpha$ being a constant of order unity
whose value depends on how one chooses to fix the gauge of time $t$.
For most calculations in this paper and in Paper~2
we choose a rather large value of the accretion rate, $\eta_s = 0.1$,
which corresponds roughly to a black hole expanding
at a tenth of the speed of light.
In astronomically realistic black holes,
where the accretion timescale is long compared to the characteristic
timescale of the black hole,
the accretion rate $\eta_s$ is unlikely to be this large.
In the calculations reported in \S\S\ref{Resultsbaryons}--\ref{Appearance}
we choose to use an unrealistically large value of $\eta_s$
partly so as to amplify the difference between the similarity solutions
and the vacuum black hole,
and partly to avoid the risk of numerical artifacts
that might possibly be associated with a tiny $\eta_s$.

The charge-augmented mass $\Mc$ at radius $r$ is defined by
\begin{equation}
\label{Mc}
  \Mc \equiv M + {Q^2 \over 2 r}
\end{equation}
where $M$ and $Q$ are the mass and charge interior to radius $r$.
In the Reissner-Nordstr\"om geometry,
the charge-augmented mass $\Mc$ is constant as a function of radius.
The mass $Q^2 / 2 r =
\int_r^\infty \left[ (Q / r^2)^2 / 8 \pi \right] 4 \pi r^2 \dd r$
on the right hand side of equation~(\ref{Mc})
is the mass-energy that would be in the electric field $Q / r^2$
outside radius $r$ if there were no charge outside $r$.
The charge-augmented mass $\Mc$ at the sonic point
is the mass that the black hole would appear to have at infinite distance
if there were no mass or charge outside $r$.
In this paper, values of radius $r$ are reported in units where
$\Mc = 1$ at the outer sonic point.
This is not a boundary condition on $\Mc$, just a choice of units.

\qmfig

The charge-to-mass $Q/\Mc$ of the black hole
at the outer sonic point
depends on the entire accretion history of the black hole.
In the similarity solutions,
this charge-to-mass $Q/\Mc$
is determined by the charge-to-mass density $q/\rho_b$
of the baryons being accreted into the black hole.
Figure~\ref{qm}
shows that
the charge-to-mass $Q/\Mc$ of the black hole
at the outer sonic point
generally increases
as the charge-to-mass density $q/\rho_b$
of baryons at the outer sonic point is increased,
although $Q/\Mc$ does turn down slightly at the largest
values of $q/\rho_b$
when the conductivity is small.
As the conductivity of the baryonic plasma is increased,
a larger charge-to-mass density $q/\rho_b$
at the sonic point is needed to produce a given
charge-to-mass $Q/\Mc$ of the black hole.
This is because a larger conductivity
allows the charge-to-mass density $q/\rho_b$
that is actually accreted through the outer horizon of the black hole
to be significantly smaller than the charge-to-mass density
at the sonic point.

\section{Results}
\label{Resultsbaryons}

This section presents results for black holes that accrete only baryons.
In Paper~2 \cite{Paper2}
the black hole is allowed to accrete dark matter in addition to baryons.

Geometric units $G = c = \Mc = 1$ are used
here and throughout the remainder of this paper,
where $\Mc$, equation~(\ref{Mc}),
is the charge-augmented interior mass of the black hole
evaluated at the outer boundary, the sonic point.

\subsection{Uncharged black hole}
\label{Schwsec}

Figure~\ref{varsSchw}
shows the simplest of the black hole solutions
considered in this paper,
that for a black hole which accretes baryons with zero charge.
Accreting no charge, the black hole is, naturally, uncharged.

\varsSchwfig

The uncharged solutions are characterized by a single free parameter,
the dimensionless accretion rate $\eta_s$
at the outer boundary, the outer sonic point.
As with all the models illustrated in this paper,
the accretion rate $\eta_s$
is set equal to $0.1$
at the outer sonic point
\begin{equation}
\label{etasonic}
  \eta_s = 0.1
\end{equation}
which roughly speaking means that the black hole
is expanding at a tenth of the speed of light.
This accretion rate is large compared to that of an astronomically realistic
black hole,
but the large value makes it easier to discern the difference in geometry
between the similarity solutions and a vacuum black hole.
Results for smaller accretion rates are qualitatively similar.
The accretion rate of the uncharged black hole
is limited to $\eta_s \lesssim 0.3616$
by the constraint $\gamma^2 - \beta^2 > 0$
at the outer sonic point.

Table~\ref{rSchw}
compares the radii $r$
of several points in the similarity solution
to those in the Schwarzschild geometry.
As can be seen, there is a considerable degree of commonality,
despite the `large' accretion rate $\eta_s = 0.1$ in the similarity solution.
In the similarity solution,
the radii are as measured in a frame falling with the baryons
(the radius $r$ is not a similarity variable,
hence the need to specify the reference frame).
In the Schwarzschild geometry,
the position of the outer sonic point
is given for a tracer relativistic fluid ($p / \rho = 1/3$)
of uncharged particles which free-fall from zero velocity at infinity.

\rSchwtab

As illustrated in Figure~\ref{varsSchw},
with no charge to repel its fall,
the baryonic fluid inside the black hole
plunges straight to a singularity at zero radius,
the infall velocity diverging as $V \sim r^{-1}$ as $r \rightarrow 0$.
As the fluid falls,
the dimensionless proper baryonic mass density
$z \equiv 4 \pi r^2 \rho_b$
tends to a constant, $z \rightarrow 0.02092$,
indicating that the baryonic density diverges as $\rho_b \sim r^{-2}$
as $r \rightarrow 0$.

The baryons develop a pressure gradient which
causes a mild outward proper acceleration $g$,
but this is not enough to prevent gravitational collapse.
It takes a proper time of $\Delta \tau = 2.164$
for the baryons to fall from the horizon to the singularity,
somewhat larger than the proper time $\Delta \tau = 4/3 \approx 1.333$
for a test particle that free falls radially from zero velocity at infinity
to fall from the horizon to the singularity of a Schwarzschild black hole.

\penroseSchwfig

Among the variables shown in Figure~\ref{varsSchw} is the homothetic scalar $H$,
equation~(\ref{H}).
Besides having the virtue of being, like the interior mass $M$,
a gauge-invariant scalar,
the homothetic scalar $H$ plays a fundamental role in describing geodesics,
see \S\ref{geodesics},
and hence in determining what things look like if you fall into a
black hole, a question addressed in \S\ref{Appearance}.
In particular,
the place where the homothetic scalar $H$ reaches a maximum
sets the location of the equivalent of the photon sphere,
which determines the angular size of the black hole perceived
by an infalling observer.
Although the boundary conditions are set at the outer sonic point,
Figure~\ref{varsSchw}
also shows a partial continuation of the similarity solution
outward from the outer sonic point.

Figure~\ref{penroseSchw}
shows a Penrose diagram of the uncharged black hole.
The Penrose diagram looks qualitatively similar to that
of the Schwarzschild solution,
except that
whereas in the Schwarzschild solution
the lower left diagonal edge of the Penrose diagram marks the antihorizon,
in the similarity solution
the same diagonal edge marks the collapse event at zero radius.
The similarity solution does not really include the collapse event itself,
and presumably the Penrose diagram
would be changed if the spacetime incorporated the collapse event.

\subsection{Black hole accreting charged, non-conducting baryons}
\label{QMsec}

Figure~\ref{varsQM}
shows a similarity solution
for a black hole which accretes charged baryons with zero conductivity.
There are two free parameters,
the accretion rate $\eta_s$,
which is set to the same value $0.1$ as before, equation~(\ref{etasonic}),
and the charge-to-mass ratio $Q/\Mc$ of the black hole,
which is set at the outer sonic point to
\begin{equation}
\label{QMsonic}
  {Q \over \Mc} = 0.8
  \ ,
\end{equation}
where $\Mc$ is the charge-augmented mass
defined by equation~(\ref{Mc}).
The condition $\gamma^2 - \beta^2 > 0$ at the outer sonic point
limits the maximum possible accretion rate and maximum charge-to-mass density
at the sonic point.
At fixed $Q/\Mc = 0.8$,
the accretion rate is limited to $\eta_s \lesssim 0.3496$.
At fixed $\eta_s = 0.1$,
the charge-to-mass ratio is limited to $Q/\Mc \lesssim 1.0060$,
the maximum value being attained at slightly less than
the maximum charge-to-mass density $q/\rho_b$,
whereat $Q/\Mc \approx 0.99979536$,
as illustrated in Figure~\ref{qm}.

\varsQMfig

The black hole charge $Q$ is taken without loss of generality to be positive.
The charge of real black holes is most likely positive, albeit minuscule,
because the larger mass-to-charge of protons compared to electrons
makes it easier for a black hole to accrete positive charge.

Unlike the uncharged black hole,
the baryons inside the charged black hole do not fall to a singularity.
Instead,
they are repelled by the charge of the black hole,
become outgoing, and drop through the Cauchy horizon.
The transition from ingoing to outgoing is marked
by the proper velocity $V$ of the similarity frame relative to the baryons
passing from $+\infty$ to $-\infty$.
As described at the end of \S\ref{inoutgoing},
the fluid remains perfectly well-behaved through this point:
the velocity $V$
passes through infinity because
the homothetic 4-vector $\xivec$ in the baryonic frame
switches from pointing forwards to backwards in time.

Table~\ref{rQM}
compares the radii $r$
of several points in the similarity solution
to those in the Reissner-Nordstr\"om (RN) geometry
with the same charge-to-mass $0.8$.
As previously found in the uncharged model of \S\ref{Schwsec},
there is a considerable degree of commonality,
despite the `large' accretion rate $\eta_s = 0.1$ in the similarity solution.
In the similarity model,
the radii are as measured in a frame falling with the baryons.
In the RN geometry,
the positions of sonic points,
and the places where charged baryons become outgoing,
and where $\gamma$ goes negative,
are given for a tracer relativistic fluid ($p / \rho = 1/3$) of particles
whose charge-to-mass is the same as that of the black hole, $0.8$,
and which free-fall from zero velocity at infinity.

\rQMtab

\shkQMfig


If the baryons could see ingoing light or matter
falling from the outside universe,
then as the baryons passed through the Cauchy horizon
they would see the light or matter infinitely blueshifted.
However,
it is being assumed here that there is no ingoing radiation or matter.
Physically, one can imagine that the baryons are sufficiently opaque
that no ingoing radiation or matter reaches the baryons near the Cauchy horizon.
In Paper~2
the effects of ingoing dark matter,
whose streaming through the outgoing baryons
leads to mass inflation near the Cauchy horizon,
will be considered explicitly.

Inside the Cauchy horizon
the baryonic fluid continues to decelerate
until the flow velocity $V$ equals the speed of sound,
at which point the proper acceleration $g$ of the fluid diverges.
Normally this is a signal that a shock must form,
which decelerates the fluid discontinuously from supersonic to subsonic.
The position of the shock is a free parameter,
which in practice is constrained over a rather narrow range of
radii where the fluid is subluminal but supersonic.
Normally the position of the shock would be fixed by
requiring that the solution downstream continue in a regular fashion.
In the case under consideration,
the gas, having been decelerated below the speed of sound by the shock,
subsequently accelerates back up to the speed of sound,
and normally one would require
that the place where the velocity accelerates back
from subsonic to supersonic be a regular sonic point,
where the acceleration is finite and preferably differentiable.
Here however the acceleration diverges for all values of the shock position,
so there is no consistent continuation of the similarity solution.

Figure~\ref{shkQM}
illustrates three possible shocks,
ranging from extremely weak (small shock velocity)
to extremely strong (highly relativistic shock velocity).
In all cases,
whereas the pre-shock fluid decelerates strongly,
the post-shock fluid re-accelerates strongly inward,
and soon accelerates back up to an irregular sonic point,
where the acceleration diverges.

The failure of the similarity solution to continue to zero radius
is discussed in \S\ref{incomplete}.

\subsection{The radial 4-gradient is timelike inside the Cauchy horizon}

An apparently small but nevertheless crucial difference in Table~\ref{rQM}
between the similarity solution and the RN geometry concerns
the radial 4-gradient $(\beta, \gamma) \equiv (\partial_t r, \partial_r r)$.
In the RN geometry the radial 4-gradient,
having switched from spacelike to timelike at the outer horizon,
then switches back to spacelike at the inner horizon.
By contrast,
in the similarity solution,
the radial 4-gradient,
having switched from spacelike to timelike
a little way inside the outer horizon,
never changes back to being spacelike.

The timelike character of the radial 4-gradient
$(\beta, \gamma)$ in the similarity solution
means that it is impossible inside the Cauchy horizon
to accelerate (with rockets, say) to a frame which is at rest in radius $r$,
with $\beta = 0$:
all locally inertial frames inside the Cauchy horizon
are falling radially inward, $\beta < 0$.
Thus a person inside the Cauchy horizon is doomed to move inward to
smaller radius $r$,
which is quite unlike the RN solution.
Because the similarity solutions do not continue to zero radius,
it is not possible to say exactly what happens,
but if $(\beta, \gamma)$ remains timelike down to zero radius,
then an infaller will hit zero radius, $r = 0$,
in a finite proper time.
The best that an infaller can do to delay the inevitable fate
is to accelerate to a frame where $\gamma = 0$,
the frame where $\beta$ is least negative.

This seems paradoxical:
if the region inside the Cauchy horizon is subluminal,
why can't an infaller stay away from zero radius?
In the context of similarity solutions,
subluminal means being able to move either outward or inward
relative to the similarity frame.
Here however
the similarity frame is itself contracting inside the Cauchy horizon.
The similarity frame moves with $r \propto t$,
and normally one would think that this means that the
similarity frame is expanding.
However,
inside the Cauchy horizon
it is consistent to think of the coordinate time $t$ as being
negative, and increasing towards $t = 0$;
alternatively, one can think of the coordinate time $t$ as being
positive and decreasing towards $t = 0$
(the ambiguity in the sign of $t$ expresses gauge freedom;
the gauge-invariant statement is that
$\eta \equiv \alpha r / t$ must be negative).
The radius $r$ of the similarity frame thus contracts to $r = 0$
at $t = 0$.
In the light of the
Penrose-Hawking singularity theorems
\cite{Penrose65,HE70,Goncalves04}
one might reasonably expect a presumably spacelike singularity
at zero radius,
although this expectation cannot be confirmed here explicitly,
because of the failure of the similarity solutions to continue
consistently to zero radius.

Parcels of baryonic fluid that fall through the Cauchy horizon later
find themselves at larger radius $r$.
Thus, again paradoxically,
even though the similarity frame is contracting from the point of
view of objects at rest in the similarity frame inside the Cauchy horizon,
the Cauchy horizon is nonetheless expanding
in the sense that baryons that fall in later find the Cauchy horizon
at larger radius $r$.

\penroseQMfig

Figure~\ref{penroseQM}
shows a tentative Penrose diagram of the spacetime under consideration.
It is tentative because it extrapolates to a spacelike central singularity,
which is not established by the similarity solution.
The Penrose diagram
presumes that the baryons
undergo a self-similar shock before falling to the central singularity.
A question mark on the diagram emphasizes the fact that
the similarity solution leaves undetermined how
the spacetime continues (or not) to the singularity.
The diagram
shows that if a self-similar shock is present inside the Cauchy horizon,
then the shock propagates inwards
from parcels of baryonic fluid accreted at later times
towards parcels of baryonic fluid accreted at earlier times.

The left (lower and upper) diagonal edges of the Penrose diagram
mark the collapse event at $r = 0$.
The similarity solution does not really include the collapse event itself,
and presumably the left part of the Penrose diagram
would be changed if the spacetime incorporated a realistic collapse event.
Notice that, according to the Penrose diagram,
a test particle that falls into the black hole and remains ingoing
will encounter a null singularity at zero radius.
It is a basic hypothesis of the similarity solution
that such ingoing particles are not actually present:
if any finite energy density of ingoing particles were present,
then that would lead to mass inflation near the inner horizon,
as considered in Paper~2.
Nevertheless,
if the similarity solution is supposed to remain valid all the
way to the collapse event at $r = 0$,
then an ingoing test particle with a zero energy-momentum tensor
would in principle approach $r = 0$,
and in so doing would pass through outgoing baryonic fluid
accreted at ever closer to the initial collapse event.

Question for the reader:
If you fall inside the Cauchy horizon,
you can hover just below the Cauchy horizon by accelerating
hard---but in which direction?
Before answering this question let us orient ourselves inside the black hole.
Before you fall into the black hole,
equip yourself with a gyroscope that is initially set
to point radially inward into the black hole.
As you fall inward,
you define the direction in which the gyroscope points
to be the immutable direction towards the black hole.
The answer to the question posed is then:
In order to hover just below the Cauchy horizon,
you must accelerate inwards towards the black hole,
in the direction indicated by the gyroscope.
You might have thought that you would have to accelerate outwards,
in the direction of baryons that fall through the Cauchy horizon
after you, but this is incorrect:
you must accelerate inwards,
in the direction of baryons that fell in before you.
Even though from your own point of view you are hovering at the Cauchy horizon,
a person who falls through the Cauchy horizon after you never sees you,
either at the Cauchy horizon or anywhere else.
On the other hand a person who falls through the Cauchy horizon before
you does see you, not necessarily close to the Cauchy horizon from their
point of view, rushing rashly by in the direction towards the black hole.
You can draw these conclusions by examining the Penrose diagram
in Figure~\ref{penroseQM}.

\velfig


It is worth emphasizing how subtle is the small difference
between the similarity and RN solutions
that leads to so dramatically different causal behavior inside the
Cauchy horizon.
Whereas in the RN solution
the radial 4-gradient
$(\beta, \gamma)$ changes from being timelike to spacelike
at the Cauchy horizon,
in the similarity solution
$(\beta, \gamma)$ remains timelike at and inside the Cauchy horizon.
Since
$2 M/r - 1 = \beta^2 - \gamma^2$,
equation~(\ref{M}),
the timelike or spacelike character of $(\beta, \gamma)$
depends on whether $M/r$ is greater or less than $1/2$.
Whereas in the RN solution
$M/r = 1/2$ at the Cauchy horizon,
in the similarity solution
$M/r = 0.510$ at the Cauchy horizon,
just slightly greater than $1/2$.
If the similarity solution is continued inside the Cauchy horizon,
then $M/r \ge 0.508$ at the terminal sonic point,
the lower limit $M/r = 0.508$
occurring for the case where there is no shock.
Unfortunately,
a proof
that $(\beta, \gamma)$ must remain timelike inside the Cauchy horizon,
along the lines of that in Appendix~\ref{proof},
fails;
but numerically $(\beta, \gamma)$ does remain timelike,
and if it does so all the way to zero radius,
then objects that fall through the Cauchy horizon
must inevitably fall to zero radius.


We have experimented with models in which the black hole charge-to-mass $Q/\Mc$
is at and near its maximum value,
but the models do not differ qualitatively from that shown in
Figure~\ref{varsQM}.
The charge-to-mass is limited by the constraint
$\gamma^2 - \beta^2 > 0$,
which is the condition,
Appendix~\ref{proof},
that the outer sonic point not be separated by
a horizon from a hypothetical asymptotically flat empty region of space
at large radius $r$.
The constraint intervenes before the black hole becomes extremal,
that is, before the inner and outer horizons of the black hole
become coincident
(which happens in the Reissner-Nordstr\"om geometry when $Q/\Mc = 1$).
There is no sign of phenomena associated with cricital collapse
\cite{Choptuik93,EC94,Gundlach03},
such as ringing, or the appearance of a naked singularity.

\subsection{Incompleteness of solutions inside the Cauchy horizons}
\label{incomplete}

In \S\ref{QMsec}
it was found that the similarity solution inside the Cauchy horizon
could not be continued consistently to zero radius,
but rather terminated at an irregular sonic point,
where the acceleration diverged.
The solution terminated whether or not a shock was introduced.
We find the same behavior
whenever the similarity solution drops inside the Cauchy horizon:
in all the cases that we have examined,
including those described in the rest of this paper,
and in \S{}IV\,B of Paper~2,
the solution terminates at an irregular sonic point,
with or without a shock,
just inside the Cauchy horizon.

Does this mean that the entire similarity solution
must be discarded as inconsistent?
The methods considered in this paper are insufficient
to supply a definitive answer to this question,
but physically it seems reasonable that the similarity solution
outside the Cauchy horizon should be valid even if there is no
consistent self-similar continuation inside the horizon.
Since no information can propagate outward from the Cauchy horizon,
the solution outside the Cauchy horizon cannot know that
the solution fails inside the Cauchy horizon.

As remarked above,
the similarity solution does not include the instant of gravitational collapse
where the black hole first forms, at $M = r = t = 0$.
But if a seed black hole, having formed with some geometry or other,
accretes in a self-similar fashion,
then it seems reasonable that the black hole could,
after many doublings of its mass, settle asymptotically
to the self-similar form.
The interior mass $M$ and charge $Q$ of the black hole are
generated self-consistently by the accretion of charged baryons,
so it seems reasonable to expect that after a sufficiently long
time the black hole would forget what happened at its formation.


The presence of any outgoing tail
of radiation produced by the gravitational collapse of the black hole
\cite{Price72,HP98a,Dafermos03,Dafermos04}
is not relevant to the question of the consistency of the similarity solutions
being considered here.
In the similarity solutions considered here,
the baryonic fluid is already outgoing,
and the presence of additional outgoing radiation cannot prevent
the baryonic fluid from passing through the outgoing Cauchy horizon.
What can prevent the baryonic fluid from dropping through the Cauchy
horizon is ingoing matter or radiation
(which leads to mass inflation; see Paper~2).
To have a causal effect on a parcel of outgoing baryonic fluid,
such ingoing matter must necessarily be accreted after the baryons.
It is thus evident that what happened to the black hole at its birth
cannot affect whether or not the baryonic fluid drops through a Cauchy horizon,
though the birth event clearly can affect what happens
in the subluminal region inside the Cauchy horizon.

Whatever the case,
the results of the present paper provide a definite set
of self-similar boundary conditions at the Cauchy horizon,
which could be supplied to a general relativistic hydrodynamic code.
Here we leave the intriguing question of what happens beyond the Cauchy horizon
to a future investigation.

\subsection{Black hole accreting charged, conducting baryons}

\velQMonefig

Figure~\ref{vel}
shows the proper velocity $V$
of the similarity frame relative to the baryonic frame
for black holes accreting baryons with a range of
electrical conductivities.
The accretion rate $\eta_s = 0.1$ and black hole charge-to-mass $Q/\Mc = 0.8$
are set to the same values as before,
equations~(\ref{etasonic}) and (\ref{QMsonic}).

The solutions reveal a surprisingly rich structure as the
dimensionless conductivity coefficient
$\kappa$, equation~(\ref{sigma}), is increased.
The baryonic fluid accreted by the black hole
undergoes one of two possible fates:
either the baryonic fluid drops through a Cauchy horizon,
or else it falls to a singularity at zero radius.
Mostly, the former fate happens if the conductivity is less
than a continuum threshold conductivity $\kappa_\infty \approx 1.304$,
while the latter fate happens if the conductivity is greater than the threshold.
However,
the baryons fall to a central singularity
not only in the continuous range of conductivities $\ge \kappa_\infty$,
but also at any of an infinite spectrum of discrete values
$\kappa_n$ ($n = 1, 2, ...$),
asymptoting to the continuum threshold $\kappa_\infty$.

Table~\ref{kappa}
lists the first four of the discrete values $\kappa_n$,
the continuum threshold value $\kappa_\infty$,
and the maximum conductivity $\kappa_{\max}$.
The maximum conductivity $\kappa_{\max}$
is set by the requirement that $\gamma^2 - \beta^2 > 0$ at the sonic point,
which is the condition that the sonic point be causally connected to
(not separated by a horizon from)
a hypothetical asymptotically flat empty region of space at large radius
(see Appendix~\ref{proof}).
Whereas the values for $\kappa_n$ and $\kappa_\infty$ in Table~\ref{kappa}
are determined numerically
(with fewer significant digits for $\kappa_\infty$ because
of the greater difficulty in locating this point precisely),
the value given for $\kappa_{\max}$ is theoretical
(inferred from the condition $\gamma^2 - \beta^2 = 0$);
numerical integration becomes unstable if
$\kappa$ is within about $0.1$ of $\kappa_{\max} \approx 67.0$.

\kappatab

Solutions between the discrete values $\kappa_n$
are characterized by the number of times $n$ that
the baryonic fluid transitions between ingoing and outgoing.
At conductivities below the first critical value, $\kappa < \kappa_1$,
the fluid transitions from ingoing to outgoing
and then drops through the Cauchy horizon.
At conductivities between the first two critical values,
$\kappa_1 < \kappa < \kappa_2$,
the fluid transitions from ingoing to outgoing
to ingoing to outgoing
before dropping through the Cauchy horizon.

Collapse to a singularity at one of the discrete conductivities $\kappa_n$
requires extreme fine-tuning:
the conductivity must be exactly equal to the critical value.
With any tiny departure from the critical conductivity,
the baryons will not fall to a singularity,
but will instead drop through the Cauchy horizon.

\varsQMkfig

Figure~\ref{velQM1}
shows two solutions
whose conductivities are just slightly above and below,
by $\mbox{a few} \times 10^{-14}$,
the first critical value $\kappa_1$.
The two solutions initially track each other closely.
The velocity $V$ passes through $\pm \infty$
as the baryons switch from ingoing to outgoing,
and then settles to a power law $V \propto r^{-1}$.
If the conductivity were exactly critical, $\kappa = \kappa_1$,
then the velocity would continue as $V \propto r^{-1}$
down to zero radius.
In the solutions shown,
the velocities depart from the power law at a small radius
$r \sim 10^{-15}$.
There is nothing special about this radius:
the departure occurs at smaller and smaller radius
as $\kappa$ approaches $\kappa_1$.
In the solution just below $\kappa_1$,
once the velocity $V$ departs from the power law,
it decreases rapidly,
and the baryons soon drop through the Cauchy horizon.
Conversely,
in the solution just above $\kappa_1$,
once the velocity $V$ departs from the power law,
it increases rapidly,
passes from $-\infty$ to $+\infty$
as the baryons transition back from outgoing to ingoing,
and settles back to another power law $V \propto r^{-1}$.
The velocity remains on this power law for some time
before steepening into another power law $V \propto r^{-4}$,
then steepening again to pass
from $+\infty$ to $-\infty$
as the baryons transition for a second time from ingoing to outgoing.
The velocity then decreases rapidly,
and the baryons finally drop through the Cauchy horizon,
at a radius some $35.4$ decades smaller than the radius
at which the below-critical solution already dropped through.

The behavior of the solutions close to the critical conductivity $\kappa_1$
is similar to that nicely described by and illustrated in Figure~1 of
Gundlach (2003) \cite{Gundlach03}
in the context of critical gravitational collapse.

Frolov and Pen (2003) \cite{FP03}
remark that the solutions just above and below
the critical solution in critical gravitational collapse
appear quite different.
They pose this as a paradox,
and conjecture that perhaps the two solutions differ only by a gauge change.
The results obtained here suggest a different interpretation,
that two spacetimes either side of a critical point
may indeed be radically different.
As the critical point is approached,
the differences, though still present, recede to ever smaller radius.


It seems possible that the oscillatory character of the solutions
with conductivities between $\kappa_1$ and $\kappa_\infty$
might perhaps be related in some way to the oscillatory character
of the curvature near the inner horizon of a rotating black hole
reported by \cite{Ori99}.

Figure~\ref{varsQMk}
illustrates a model with conductivity
$\kappa = 0.8$
below the first critical threshold $\kappa_1$.
Like the zero conductivity model shown in Figure~\ref{varsQM},
the baryonic fluid ends up dropping through the Cauchy horizon.
The finite conductivity allows currents to flow
so that, as Figure~\ref{varsQMk} shows,
the interior charge $Q$ decreases along the path of the infalling baryons,
whereas in the non-conducting case the interior charge $Q$ would be constant.
The conductivity is however not so large as to allow the baryonic fluid
to neutralize itself entirely, so $Q$ remains positive,
and, like the non-conducting case shown in Figure~\ref{varsQM},
the baryons end up dropping through the Cauchy horizon.

In the conducting model shown in Figure~\ref{varsQMk}
the dimensionless charge density $z_q$
is negative inside a certain radius,
whereas in the non-conducting model of Figure~\ref{varsQM},
the charge density remained everywhere positive.
How can the interior charge $Q$ be positive if the charge density
inside a certain radius is everywhere negative?
Because conservation of charge in the similarity frame
depends on the current $j$ as well as the charge density $q$.


\varsQMcfig

\scalingQMcfig

\subsection{Discrete self-similarity at threshold}

The self-similar solution at the continuum threshold conductivity
$\kappa_\infty$
is of special interest.
The solution displays a discrete self-similarity strikingly
reminiscent of the discrete self-similarity in critical gravitational
collapse first discovered numerically by Choptuik (1993) \cite{Choptuik93};
see \cite{Gundlach03} for a review.

Figure~\ref{varsQMc}
shows several variables in the similarity solution
at the continuum threshold $\kappa_\infty$
(or at least as close to the continuum threshold as we could discern,
numerically).
All the variables oscillate in phase,
the solution at each period appearing to be
a scaled version of the solution at any other period.

Figure~\ref{scalingQMc}
shows the velocity $V$ scaled by $r^{1.7081}$.
The scaled velocity plotted in Figure~\ref{scalingQMc}
passes through 31 cycles, each $3.9463$ decades wide in radius,
covering in all about 120 decades of radius.

The discrete self-similarity is not exact.
For example, the first cycle, which starts outside the horizon,
starts out noticeably different in shape from the later cycles.
Presumably the discrete self-similarity operates asymptotically,
the shape of the cycle converging asymptotically to a certain form
as the number of cycles increases.

Koike, Hara \& Adachi (1995) \cite{KHA95}
explained the discrete self-similarity observed in gravitational collapse
as resulting from a dominant unstable mode of perturbation
to an unperturbed self-similar solution.
This idea has been explored and built upon by several authors
\cite{HE95,Maison96,KHA99,HM01,BCGN02};
see \cite{Harada03} for a review.
Possibly a similar explanation could apply in the present case.

\appearfig

\section{Appearance of the black hole}
\label{Appearance}

What does it actually look like if you fall inside
one of the black holes described in this paper?
A full answer to this fascinating question is beyond our scope,
but one can begin to get an idea by answering a simpler question:
how big does the black hole appear on the sky as you fall into it?
Or at least, how big would the black hole appear to you
if you could actually see it through the fog of accreting baryons,
an ominous black disk silhouetted against the starry universe outside?

It would of course have to be you, not a remote instrument,
that observed the inside of the black hole,
since an instrument inside the horizon
would not be able to relay its observations
to you outside the horizon.
You should be sure to fall into a supermassive black hole,
since you could easily survive the tidal forces well inside the black hole,
whereas a stellar-massed black hole would dismember you
long before you reached the horizon.
A useful rule to remember is that, for an observer in circular orbit,
the tidal force is approximately one gee per meter
at a radius where the orbital period is one second,
a fact which the interested reader may care to confirm.

In a Reissner-Nordstr\"om black hole
the perceived edge of the black hole
is set by photons orbiting at the photon sphere.
Such photons fall from the outside universe
into the unstable orbit at the photon sphere,
circulate the black hole an arbitrarily large number of times,
and then either escape back out to infinity or else fall into the black hole.

In the self-similar black holes of the present paper,
there is no photon sphere as such,
no place where photons can orbit in circles for ever.
There is however a separatrix between photons from the outside universe
that do or do not fall into the black hole,
and it is this separatrix,
which we call the photon sphere equivalent,
that defines the perceived edge of the black hole as viewed on the sky.

Figure~\ref{appear} shows, for the three models illustrated in
Figures~\ref{varsSchw}, \ref{varsQM}, and \ref{varsQMc},
the angular size $\chi_\textrm{ph}$
and blueshift of photons from the edge of the black hole,
as observed either in the baryonic rest frame,
or in a rest frame that free-falls radially from zero velocity at infinity
(in the latter case,
the frame is the same as that of the pressureless dark matter
considered in Paper~2).
The two points of view are related by a radial Lorentz boost.
The observed angular size $\chi_\textrm{ph}$ of the black hole
(the subscript ph signifying photons from the photon sphere equivalent)
is given by equation~(\ref{chiph}),
and the observed blueshift of photons at the edge of the black hole
is given by equation~(\ref{blueph}).

The appearance of an uncharged, non-conducting black hole,
shown in the left panel of Figure~\ref{appear},
is similar to that of the corresponding vacuum black hole,
the Schwarzschild solution.
As the observer plunges to the central singularity at zero radius,
the black hole increases monotonically in angular size,
reaching $90^\circ$ at the singularity.
The blueshift of photons at the edge of the black hole tends to infinity.
The appearance can be attributed to the enormous tidal force near
the singularity.
The same tidal force that
stretches the infaller radially and crushes them horizontally
also aberrates photons so that they appear to come from a thin horizontal
blueshifted band of light on the sky.
Photons from above the observer are redshifted,
while photons from the thin horizontal band are highly blueshifted.
Although the blueshift at the edge of the black hole
tends to infinity as the observer approaches the singularity,
the amount of time that the observer sees pass by in the outside universe,
the integral of blueshift over proper time, is finite.

The appearance of a charged, non-conducting black hole,
shown in the middle panel of Figure~\ref{appear},
is again similar to that of its vacuum counterpart,
the Reissner-Nordstr\"om (RN) solution,
down to the inner horizon.
From the point of view of an outgoing observer,
such as one in the rest frame of the charged baryonic fluid,
the black hole increases in angular size until, at the Cauchy horizon,
the black hole covers the entire sky.
The view of the outside universe correspondingly decreases in size
and becomes brighter and more blueshifted,
until the view disappears at the Cauchy horizon
in an infinitely bright, blueshifted, concentrated flash.
The entire future of the outside universe
passes in that infinitely blueshifted flash.
It should be commented that such an illuminating experience
could not actually happen in reality,
because photons from the outside universe are necessarily ingoing,
and the presence of even the tiniest amount of ingoing matter or radiation
will prevent passage through the Cauchy horizon---see \S{}V\,A of Paper~2.

From the point of view of an ingoing observer on the other hand,
such as one who free-falls radially from zero velocity at infinity,
the charged, non-conducting black hole
(middle panel of Figure~\ref{appear}, again)
first increases in angular size,
then shrinks as the observer approaches the inner horizon.
Just above the inner horizon, the angular size of the black hole is finite,
and the blueshift is still finite,
just as it is in the RN geometry.
In the RN solution,
an ingoing observer would pass through the inner horizon,
and in the instant of passage
would see a new image of the outside universe,
reflected by the gravitationally repulsive core of the black hole,
appear at the center of the black hole in an infinitely blueshifted flash
containing the entire history of the universe.
In the similarity solution by contrast,
the ingoing observer does not pass through the inner horizon,
and no new image of the universe appears in an infinitely blueshifted flash.
Instead,
the ingoing observer falls to zero radius while still remaining
just outside the inner horizon.
As the ingoing observer falls to smaller radius,
they encounter outgoing baryons accreted at ever earlier times,
back to the initial collapse event at $r = 0$.
In this regime where the ingoing observer remains close to the inner horizon,
the appearance of the outside universe
evolves simply by becoming more and more Lorentz-boosted.
Although the blueshift tends to infinity as the ingoing observer approaches
zero radius,
the amount of time that the ingoing observer sees
pass by in the outside universe remains finite.
This is unlike the outgoing observer,
who (if they could actually see it)
would see the entire future of the universe pass by
as they cross the Cauchy horizon.

The right panel of Figure~\ref{appear} shows
the appearance of the discretely self-similar charged baryonic model,
with conductivity equal to the continuum threshold conductivity
$\kappa = \kappa_\infty$,
from Figure~\protect\ref{varsQMc}.
As is evident from Figure~\ref{appear},
the oscillations in the model
are mildy visible to a free-fall observer,
but, curiously,
practically invisible to an observer at rest in the baryonic frame.
In the baryonic frame,
the apparent angular radius of the black hole remains near $90^\circ$
all the way from the outer sonic point to the singularity at zero radius.
Near the singularity, the $90^\circ$ size can be attributed to the enormous
tidal force, as in the Schwarzschild geometry,
but at large radii the near $90^\circ$ size is presumably fortuitous,
a happenstance of the relativistic aberration
from the motion of baryons at and inward of the sonic point.
As in the case of the uncharged black hole,
although the blueshift at the edge of the black hole
tends to infinity as the observer approaches the singularity,
the amount of time that the observer sees pass by in the outside universe
remains finite.

For those interested in carrying out approximate ray-tracing
inside one of the black holes described in this paper,
it is useful to note that the function $H(X)$,
which plays the essential part in ray-tracing,
equation~(\ref{phi}),
is reasonably approximated as a cubic polynomial in $X$
if the baryons collapse straight to a central singularity,
or as a quartic polynomial in $X$
if the baryons drop through the Cauchy horizon.
The approximation becomes exact in the asymptotic limit of
small accretion rates and small conductivities,
where $H(X)$,
equation~(\ref{XHlimit}),
becomes a cubic polynomial if the black hole charge is zero
(Schwarzschild limit),
or a quartic polynomial if the black hole charge is non-zero
(Reissner-Nordstr\"om limit).
Usefully,
the approximation of $H(X)$ as a cubic or quartic polynomial
is not bad even at finite accretion rates and conductivities.

\section{Summary}
\label{Summary}

In this paper we have investigated self-similar solutions
for spherically symmetric charged black holes
that accrete a relativistic fluid ($p_b / \rho_b = 1/3$) of charged,
electrically conducting baryons.
The solutions are characterized by three free parameters:
the accretion rate,
the charge-to-mass of the black hole
(generated self-consistently by the charge-to-mass ratio of accreted baryons),
and the electrical conductivity.
We do not require that solutions be regular
at zero radius,
but rather integrate inwards from boundary conditions established
at a regular sonic point outside the outer horizon of the black hole.

The accreted charged baryons undergo one of two possible fates:
either they plunge directly to a spacelike singularity at zero radius,
or else they drop through a Cauchy horizon.
In the latter case the
baryons probably undergo a shock just inside the Cauchy horizon,
but, whether or not a shock occurs,
the similarity solutions do not continue consistently to zero radius,
but rather terminate at an irregular sonic point
where the proper acceleration diverges.
We argued in \S\ref{incomplete}
that the failure of the similarity solutions to continue to zero radius
inside the Cauchy horizon does not invalidate the solutions as a whole,
because the failure is hidden behind the Cauchy horizon,
and cannot be communicated to the solution outside the Cauchy horizon.
It remains a matter for future numerical investigation to discover
what really happens beyond the Cauchy horizon.

In the solutions where the baryons drop through the Cauchy horizon,
the geometry inside the Cauchy horizon differs in one crucial respect
from the corresponding vacuum solution, the Reissner-Nordstr\"om (RN) geometry.
Whereas in the RN geometry the radial 4-gradient is spacelike
inside the Cauchy horizon,
in the similarity solutions the radial 4-gradient is timelike.
This means that in the similarity solution it is impossible to remain at
rest (at fixed radius $r$) inside the Cauchy horizon:
all locally inertial frames necessarily fall to smaller radii.
Because the similarity solutions do not continue consistently to zero radius,
we cannot say exactly what happens,
but if the radial gradient remains spacelike all the way to zero radius,
then the baryons will fall to zero radius,
where there is presumably a spacelike singularity.

It may seem paradoxical that the spacetime inside the Cauchy horizon
is subluminal, yet nevertheless the baryonic fluid must flow to smaller radii.
The resolution of this paradox is that
from the point of view of observers inside the Cauchy horizon,
the Cauchy horizon is actually contracting to smaller radii.
Notwithstanding this contraction,
observers who fall through the Cauchy horizon at later times
find the Cauchy horizon at larger radii.
A consistent way to understand this is
to think of coordinate time as running backwards inside the Cauchy horizon
(though of course proper time always runs forward in the usual way).
The topsy-turvy world inside the Cauchy horizon
was discussed in \S\ref{QMsec}.

The question of whether the baryons plunge to a singularity
or drop through the Cauchy horizon
depends on the parameters, as summarized immediately below.

If the black hole is uncharged
(because the baryons it accretes are uncharged),
then the baryons plunge to a central spacelike singularity,
as in the Schwarzschild solution.

If the black hole is charged and the conductivity is zero,
then the accreted baryons,
repelled by the charge of the black hole
generated by previously accreted baryons,
naturally become outgoing,
and drop through the outgoing inner horizon, the Cauchy horizon.

If the black hole is charged
and if the dimensionless electrically conductivity $\kappa$
exceeds a certain threshold value $\kappa_\infty$
(whose value depends on the accretion rate and charge-to-mass ratio),
then the baryonic fluid can neutralize inside the black hole,
and plunges to a spacelike singularity.
If on the other hand the conductivity is less than the threshold value
$\kappa_\infty$,
then in most cases the baryons drop through the Cauchy horizon.
However,
there is also a discrete spectrum $\kappa_n$ ($n = 1, 2, ...$)
of conductivities,
asymptoting to the threshold $\kappa_\infty$,
at which the baryons plunge to a central singularity
rather than dropping through the Cauchy horizon.
Solutions between the discrete values $\kappa_n$
are characterized by the number of times that the baryonic fluid
oscillates between ingoing and outgoing
before dropping through the Cauchy horizon.

At the threshold conductivity $\kappa_\infty$,
the baryonic fluid oscillates between ingoing and outgoing an
infinite number of times before hitting the central singularity,
and the solution exhibits a discrete self-similar behavior
reminiscent of that observed in critical collapse
\cite{Choptuik93,EC94,Gundlach03}.

In \S\ref{Appearance}
we discussed what an observer who falls inside one of the black holes
considered in this paper would see.

Mass inflation at the Cauchy horizon
\cite{PI90,Dafermos04}
does not occur in the solutions considered in this paper.
Mass inflation requires the presence of both ingoing and outgoing fluids
near the inner horizon,
and in this paper we have deliberately chosen to consider only
a single baryonic fluid.
Although a finite conductivity does allow oppositely charged fluids
to diffuse through each other,
this is not enough to permit mass inflation.
In a companion paper, Paper~2 \cite{Paper2},
we allow the black hole to accrete not only baryons but also dark matter,
and it will be seen that streaming between baryons and dark matter
leads to mass inflation.

\begin{acknowledgements}
We thank Steven Carlip for helpful conversations.
This work was supported in part by
NSF award ESI-0337286.
\end{acknowledgements}

\onecolumngrid

\appendix

\section{General Spherically Symmetric Tetrad}
\label{generaltetrad}

This appendix presents expressions
for the vierbein,
metric,
connection coefficients,
and Einstein tensor,
for the most general spherically symmetric orthonormal tetrad.

It is convenient to proceed in two steps,
carried out in \S\S\ref{arbitraryr} and \ref{general}
immediately below.
The first step is to generalize the vierbein adopted in the text,
equations~(\ref{vierbein}),
to the case of arbitrary radial coordinate.
The second step is to complete the generalization.

\subsection{Arbitrary radial coordinate}
\label{arbitraryr}

In the text,
the gauge choices
$\nu = \mu = 0$
and
$\lambda = 1$,
equations~(\ref{gauge}),
were imposed on the general vierbein~(\ref{genvierbein}),
bringing the vierbein to the form~(\ref{vierbein}).
The gauge choice $\lambda = 1$
is equivalent to setting the radial coordinate $r$
to be the circumferential radius,
defined such that the proper circumference of a circle is $2 \pi r$.
In this subsection of Appendix~\ref{generaltetrad},
the gauge choices
$\nu = \mu = 0$
are retained,
but $\lambda$ is permitted to be arbitrary,
so that the radial coordinate is arbitrary.

Denote the arbitrary radial coordinate $\rrarb$
and the corresponding vierbein coefficients
$\alphararb$, $\betararb$, $\gammararb$, and $\lambdararb$
with primes,
to distinguish them from the circumferential radial coordinate $\rnorm$
and vierbein coefficients
$\alphanorm$, $\betanorm$, $\gammanorm$
considered in the main text.
The vierbein
${e_m}^\mu$ are then
\begin{equation}
\label{vierbeine}
  {e_0}^0 = \alphararb
  \ , \quad
  {e_i}^0 = 0
  \ , \quad
  {e_0}^i = \betararb \, \hat x_i
  \ , \quad
  {e_i}^j
  =
  \gammararb \, \hat x_i \hat x_j
  + \lambdararb \, (\delta_{ij} - \hat x_i \hat x_j)
\end{equation}
with inverse ${e^m}_\mu$
\begin{equation}
  {e^0}_0
  =
  {1 \over \alphararb}
  \ , \quad
  {e^0}_i
  =
  0
  \ , \quad
  {e^i}_0
  =
  - {\betararb \over \alphararb \gammararb} \, \hat x_i
  \ , \quad
  {e^j}_i
  =
  {1 \over \gammararb} \, \hat x_i \hat x_j
  + {1 \over \lambdararb} \, (\delta_{ij} - \hat x_i \hat x_j)
  \ .
\end{equation}
The metric is
\begin{equation}
\label{metrice}
  \dd s^2
  =
  - \, {\dd t^2 \over \alphararb^2}
  + {1 \over \gammararb^2} \left(
  \dd \rrarb
  - {\betararb \dd t \over \alphararb}
  \right)^2
  + {\rrarb^2 \dd o^2 \over \lambdararb^2}
  \ .
\end{equation}
It is evident from the metric~(\ref{metrice}) that the quantity
$\rnorm$ defined by
\begin{equation}
\label{rnorm}
  \rnorm \equiv {\rrarb \over \lambdararb}
\end{equation}
is the circumferential radius $r$ of the main text.
A relation between the new vierbein coefficients
$\alphararb$, $\betararb$, $\gammararb$, $\lambdararb$
of equation~(\ref{vierbeine})
and the vierbein coefficients
$\alphanorm$, $\betanorm$, $\gammanorm$
in the main text is most easily established by noting that
the directed derivatives $\partial_m$, equation~(\ref{directedderivative}),
are independent of the choice of coordinate system $x^\mu$,
and in particular are independent of the choice of radial coordinate.
Thus the directed derivatives in the time $t$
and radial $r$ directions must be the same
irrespective of the choice of radial coordinate, $\rnorm$ or $\rrarb$:
\begin{eqnarray}
\label{directedderivativetre}
&&
  \partial_t
  =
  \alphanorm \left. {\partial \over \partial t} \right|_{\rnorm}
  + \betanorm \left. {\partial \over \partial \rnorm} \right|_t
  =
  \partialrarb_t
  =
  \alphararb \left. {\partial \over \partial t} \right|_\rrarb
  + \betararb \left. {\partial \over \partial \rrarb} \right|_t
\nonumber
\\
&&
  \partial_r
  =
  \left. \gammanorm {\partial \over \partial \rnorm} \right|_t
  =
  \partialrarb_r
  =
  \left. \gammararb {\partial \over \partial \rrarb} \right|_t
  \ .
\end{eqnarray}
It follows from equalities~(\ref{directedderivativetre}) that
$\alphanorm$, $\betanorm$, $\gammanorm$
in the main text are related to
$\alphararb$, $\betararb$, $\gammararb$, $\lambdararb$
by
\begin{equation}
\label{alphanorm}
  \alphanorm \equiv \alphararb
  \ , \quad
  \betanorm
  \equiv
  {\betararb \over \lambdararb}
  + \rrarb \, \partial_t \left( {1 \over \lambdararb} \right)
  \ , \quad
  \gammanorm
  \equiv
  {\gammararb \over \lambdararb}
  + \rrarb \, \partial_r \left( {1 \over \lambdararb} \right)
  \ .
\end{equation}
The inverse formulae for $\betararb$ and $\gammararb$
in terms of $\betanorm$ and $\gammanorm$
are
$\betararb = \betanorm \lambdararb + \rnorm \, \partial_t \lambdararb$
and
$\gammararb = \gammanorm \lambdararb + \rnorm \, \partial_r \lambdararb$.
All the formulae given in the main text now carry through
with $r$, $\alpha$, $\beta$, and $\gamma$
given by equations~(\ref{rnorm}) and (\ref{alphanorm}).
One useful relation to note is that
the proper acceleration $\grarb$ and proper velocity gradient $\hrarb$
defined analogously to $g$ and $h$ of equations~(\ref{g}) and (\ref{h}),
\begin{equation}
  \grarb
  \equiv
  - \, \partial_r \ln\alphararb
  \ , \quad
  \hrarb
  \equiv
  {\partial \betararb \over \partial \rrarb}
  - \partial_t \ln\gammararb
  - \betararb \, {\partial \ln\alphararb \over \partial \rrarb}
\end{equation}
are the same as their unprimed counterparts
\begin{equation}
  g = \grarb
  \ , \quad
  h = \hrarb
  \ .
\end{equation}

\subsection{General case}
\label{general}

The most general form of the vierbein
${e_m}^\mu$
consistent with spherical symmetry
is given \cite{Robertson} by equation~(\ref{genvierbein}).
The general vierbein can be obtained from the vierbein
of the previous subsection, equation~(\ref{vierbeine}),
by boosting the tetrad frame radially
by boost angle $\xi$,
and rotating the tetrad frame about the radial direction
by angle $\zeta$.
The boosted and rotated vierbein coefficients
(doubly primed,
to distinguish them from the singly primed coefficients
of the previous subsection)
are then
\begin{equation}
\label{genvierbeinp}
  \alphagen = \alphararb \cosh \xi
  \ , \ \ 
  \nugen = \alphararb \sinh \xi
  \ , \ \ 
  \betagen = \betararb \cosh \xi + \gammararb \sinh \xi
  \ , \ \ 
  \gammagen = \gammararb \cosh \xi + \betararb \sinh \xi
  \ , \ \ 
  \lambdagen = \lambdararb \cos \zeta
  \ , \ \ 
  \mugen = \lambdararb \sin \zeta
  \ .
\end{equation}
Equations~(\ref{genvierbeinp}) imply the relations
\begin{equation}
  \alphararb^2
  =
  \alphagen^2 - \nugen^2
  \ , \quad
  \gammararb^2 - \betararb^2
  =
  \gammagen^2 - \betagen^2
  \ , \quad
  \alphararb \gammararb
  =
  \alphagen \gammagen - \betagen \nugen
  \ , \quad
  \alphararb \betararb
  =
  \alphagen \betagen - \gammagen \nugen
  \ , \quad
  \lambdararb^2
  =
  \lambdagen^2 + \mugen^2
\end{equation}
which are sufficient to determine
the singly primed, unboosted, unrotated vierbein coefficients
in terms of the doubly primed, boosted, rotated vierbein coefficients.
The vierbein
${e_m}^\mu$
are
[the following is the same as equation~(\ref{genvierbein}),
but with doubly primed coefficients]
\begin{eqnarray}
&&
  {e_0}^0 = \alphagen
  \ , \quad
  {e_i}^0 = \nugen \, \hat x_i
  \ , \quad
  {e_0}^i = \betagen \, \hat x_i
  \ ,
\nonumber
\\
&&
  {e_i}^j
  =
  \gammagen \, \hat x_i \hat x_j
  + \lambdagen \, (\delta_{ij} - \hat x_i \hat x_j)
  + \mugen \, \varepsilon_{ijk} \hat x_k
\end{eqnarray}
with inverse
${e^m}_\mu$
\begin{eqnarray}
&&
  {e^0}_0
  =
  {\gammagen \over \alphagen \gammagen - \betagen \nugen}
  \ , \quad
  {e^0}_i
  =
  - \,
  {\nugen \over \alphagen \gammagen - \betagen \nugen}
  \, \hat x_i
  \ , \quad
  {e^i}_0
  =
  - \,
  {\betagen \over \alphagen \gammagen - \betagen \nugen}
  \, \hat x_i
  \ ,
\nonumber
\\
&&
  {e^j}_i
  =
  {\alphagen \over \alphagen \gammagen - \betagen \nugen}
  \, \hat x_i \hat x_j
  + {\lambdagen \over \lambdagen^2 + \mugen^2}
  \, (\delta_{ij} - \hat x_i \hat x_j)
  - {\mugen \over \lambdagen^2 + \mugen^2}
  \, \varepsilon_{ijk} \hat x_k
  \ .
\end{eqnarray}
Boosting and rotating the tetrad has no effect on the coordinates $x^\mu$,
and therefore no effect on the metric,
which continues to be given in terms of
$\alphararb$, $\betararb$, $\gammararb$, and $\lambdararb$
by equation~(\ref{metrice}).
In terms of the boosted and rotated vierbein coefficients
$\alphagen$, $\nugen$,
$\betagen$, $\gammagen$,
$\lambdagen$, and $\mugen$,
the metric is
\begin{equation}
  \dd s^2
  =
  - \, {\dd t^2 \over \alphagen^2 - \nugen^2}
  + {\alphagen^2 - \nugen^2
  \over (\alphagen \gammagen - \betagen \nugen)^2}
  \left[
    \dd \rprime
    - {(\alphagen \betagen - \gammagen \nugen) \dd t
    \over \alphagen^2 - \nugen^2}
  \right]^2
  + {\rprime^2 \dd o^2 \over \lambdagen^2 + \mugen^2}
\end{equation}
with determinant
\begin{equation}
  g
  =
  -
  \left[
  {\rprime^2 \over
  \left( \alphagen \gammagen - \betagen \nugen \right)
  \left( \lambdagen^2 + \mugen^2 \right)}
  \right]^2
  \ .
\end{equation}
The directed derivatives
$\partialgen_t$ and $\partialgen_r$
in the time $t$ and radial $r$ directions are
Lorentz boosted versions of the directed derivatives
$\partial_t$ and $\partial_r$ of the previous section,
equations~(\ref{directedderivativetre}):
\begin{subequations}
\begin{eqnarray}
  \partialgen_t
  &=&
  \alphagen {\partial \over \partial t}
  + \betagen {\partial \over \partial \rprime}
  \ = \ 
  \cosh \xi \, \partial_t + \sinh \xi \, \partial_r
\\
  \partialgen_r
  &=&
  \nugen {\partial \over \partial t}
  + \gammagen {\partial \over \partial \rprime}
  \ = \ 
  \sinh \xi \, \partial_t + \cosh \xi \, \partial_r
  \ .
\end{eqnarray}
\end{subequations}
In terms of doubly primed vierbein coefficients,
the tetrad connection coefficients are explicitly
\begin{subequations}
\begin{eqnarray}
  \Gamma_{i00}
  &=&
  \hat x_i
  \left[
    {\partial \nugen \over \partial t}
    +
    {\partial \gammagen \over \partial \rprime}
    -
    \partialgen_r \ln \left(
      \alphagen \gammagen - \betagen \nugen
    \right)
  \right]
\\
  \Gamma_{i0j}
  &=&
  \hat x_i \hat x_j
  \left[
    {\partial \alphagen \over \partial t}
    +
    {\partial \betagen \over \partial \rprime}
    -
    \partialgen_t \ln \left(
      \alphagen \gammagen - \betagen \nugen
    \right)
  \right]
  +
  \left( \delta_{ij} - \hat x_i \hat x_j \right)
  \left[
    {\betagen \over \rprime}
    -
    {1 \over 2} \,
    \partialgen_t \ln \left( \lambdagen^2 + \mugen^2 \right)
  \right]
\\
  \Gamma_{ij0}
  &=&
  - \, \varepsilon_{ijk} \, \hat x_k \,
  {\lambdagen \mugen \over \lambdagen^2
  + \mugen^2} \,
  \partialgen_t \ln \left( \mugen \over \lambdagen \right)
\\
  \Gamma_{ijk}
  &=&
  \left( \delta_{ik} \hat x_j - \delta_{jk} \hat x_i \right)
  \left[
    {\gammagen - \lambdagen \over \rprime}
    -
    {1 \over 2} \,
    \partialgen_r \ln \left( \lambdagen^2 + \mugen^2 \right)
  \right]
  - \varepsilon_{ijl} \left[
    \hat x_k \hat x_l \,
    {\lambdagen \mugen
    \over \lambdagen^2 + \mugen^2} \,
    \partialgen_r \ln \left( {\mugen \over \lambdagen} \right)
    +
    \left( \delta_{kl} - \hat x_k \hat x_l \right)
    {\mugen \over \rprime}
  \right]
  \ .
\nonumber
\\
\end{eqnarray}
\end{subequations}
Tensor quanties follow straightforwardly
by boosting and rotating the tetrad frame from the
$\nu = 0$, $\mu = 0$
frame of the previous subsection, \S\ref{arbitraryr}.
For example,
the Einstein tensor is
\begin{equation}
  G^{mn}
  =
  2 \left[
    \left( R + S \right) \hat t^m \hat t^n
    + F \left( \hat t^m \hat r^n + \hat r^m \hat t^n \right)
    + P \, \hat r^m \hat r^n
    + S \left( \eta^{mn} - \hat r^m \hat r^n \right)
  \right]
\end{equation}
where $F$, $R$, $P$, $S$
are given by equations~(\ref{FRPS})
[with $r$, $\alpha$, $\beta$, and $\gamma$
being unprimed quantities,
given in terms of primed quantities
by equations~(\ref{rnorm}) and (\ref{alphanorm}),
and thence in terms of doubly primed quantities
by the inverse of equations~(\ref{genvierbeinp})],
and the 4-vectors $\hat t^m$ and $\hat r^n$
are respectively the unit time vector and the unit radial vector
boosted by boost angle $\xi$ out of the
$\nu = 0$, $\mu = 0$ frame
\begin{eqnarray}
  &&
  \hat t^0 = \cosh \xi
  \ , \quad
  \hat t^i = - \, \sinh \xi \, \hat x_i
  \ ,
\\
  &&
  \hat r^0 = - \, \sinh \xi
  \ , \quad
  \hat r^i = \cosh \xi \, \hat x_i
  \ .
\end{eqnarray}
Note that the rotation of the tetrad frame
by angle $\zeta$ about the radial direction
leaves the Einstein tensor $G^{mn}$,
and the energy-momentum tensor $T^{mn}$,
and more generally any symmetric tensor of rank 2, unchanged.

\section{Proof that $\gamma^2 - \beta^2 > 0$ at the outer sonic point}
\label{proof}

\twocolumngrid

This Appendix proves, rather unprettily,
the assertion in \S\ref{boundaryconditions} that,
for the sonic point at the outer boundary to be causally connected to
(not separated by a horizon from)
an asymptotically flat empty region of space at large radius $r$,
it is necessary that the radial 4-gradient $(\beta, \gamma)$ be spacelike
at the sonic point,
$\gamma^2 - \beta^2 > 0$.
The proof applies to the self-similar solutions not only of the present paper,
but also of Paper~2,
where the black hole accretes dark matter in addition to baryons.
Quantities below subscripted $d$ refer to dark matter.

The assertion will be proven for the case where
the geometry consists of a spherically symetric self-similar region
surrounded by empty space.
As remarked in \S\ref{boundaryconditions},
the self-similar solutions modeled in this paper
do not necessarily continue self-consistently to infinite radius.
We suppose therefore that the self-similar region is truncated at some
point outside the outer sonic point,
and that space beyond this truncation point is empty.
Actually the proof works
if the region outside the self-similar region is not empty;
it suffices to assume that
$\gamma^2 - \beta^2 > 0$
at the outermost point to which the the self-similar solution
is considered to extend.

According to the definition of interior mass,
$\gamma^2 - \beta^2 = 1 - 2 M / r$,
equation~(\ref{M}).
In the hypothetical empty portion of space at large $r$,
the geometry is necessarily Reissner-Nordstr\"om.
By assumption, the empty region is outside any horizon,
which implies that
$\gamma^2 - \beta^2 > 0$
in the empty region.

Now consider the self-similar region of space.
The differential equation governing the self-similar evolution of
$\gamma^2 - \beta^2$
is, in the baryonic frame,
\begin{equation}
\label{dgb}
  {\dd ( \gamma^2 - \beta^2 ) \over \dd x}
  =
  2 \xi^r
  \left[ ( 1 {+} w ) z \beta \gamma + z_d \beta_d \gamma_d \right]
\end{equation}
which includes a contribution from pressureless dark matter (Paper~2).
To prove that $\gamma^2 - \beta^2$ remains positive,
assume the opposite,
that there is a point where $\gamma^2 - \beta^2$
drops through zero from positive to negative.
At the point where $\gamma^2 - \beta^2$ passes through zero,
the derivative on the right hand side of equation~(\ref{dgb}) must be negative,
so consider the various terms in this expression.
As remarked in \S\ref{inoutgoing},
consistency of the self-similar solutions requires that
$\xi^r$ always remain of the same positive sign,
since if $\xi^r$ changed sign it would indicate that the baryonic fluid turns
back on itself, which physically cannot happen.
The dimensionless proper baryonic and dark matter densities $z$ and $z_d$
in equation~(\ref{dgb}) are necessarily positive.
By assumption,
the dark matter is causally connected to, and indeed falls from,
the outer empty region of space, where $\gamma_d > 0$.
Since $\gamma_d^2 - \beta_d^2 = \gamma^2 - \beta^2$,
it follows that $\gamma_d > 0$
throughout the region where
$\gamma^2 - \beta^2 > 0$.
Since the baryonic 4-vector $(\beta, \gamma)$
is related to the dark matter 4-vector $(\beta_d, \gamma_d)$
by a Lorentz transformation,
equation~(5) of Paper~2,
it follows likewise that $\gamma > 0$
throughout the region where
$\gamma^2 - \beta^2 > 0$.
For the derivative on the right hand side of equation~(\ref{dgb})
to be negative thus requires that
at least one of $\beta$ or $\beta_d$ be negative.
Suppose without loss of generality that it is $\beta$ that is negative.
If $\beta < 0$,
then
$\gamma - \beta > \gamma + \beta$,
and for $\gamma^2 - \beta^2$ to pass through zero,
it is necessary that $\gamma + \beta = 0$.
Now
$\gamma + \beta = \gamma ( 1 - V ) + \eta$
according to equation~(\ref{V}) for the velocity $V$.
But $\gamma \ge 0$, and $V < 1$ since the point is outside the horizon,
so
$\gamma + \beta \ge \eta$.
But
$\eta = 1 / \xi^t$ is a strictly positive quantity
everywhere outside the horizon.
Thus $\gamma + \beta$ is strictly positive,
contradicting the proposition that $\gamma + \beta = 0$.
This proves the theorem.

\end{document}